\documentclass[11pt,oneside,english,reqno]{amsart}
\usepackage[T1]{fontenc}
\usepackage[latin9]{inputenc}
\usepackage{verbatim}
\usepackage{amsthm}
\usepackage{amsbsy}
\usepackage{amstext}
\usepackage{amssymb}
\usepackage{esint}
\usepackage{xargs}[2008/03/08]

\makeatletter
\numberwithin{equation}{section}
\numberwithin{figure}{section}
\theoremstyle{plain}
\newtheorem{thm}{\protect\theoremname}
  \theoremstyle{remark}
  \newtheorem{rem}[thm]{\protect\remarkname}
  \theoremstyle{definition}
  \newtheorem{example}[thm]{\protect\examplename}

\usepackage{mathpazo}
\newcommand{\ie}{\textit{i.e.}}
\newcommand{\eg}{\textit{e.g.}}
\usepackage{color}

\makeatother

\usepackage{babel}
  \providecommand{\examplename}{Example}
  \providecommand{\remarkname}{Remark}
\providecommand{\theoremname}{Theorem}

\begin{document}

\global\long\def\ga{\alpha}
\global\long\def\gb{\beta}
\global\long\def\ggm{\gamma}
\global\long\def\go{\omega}
\global\long\def\ge{\epsilon}
\global\long\def\gs{\sigma}
\global\long\def\gd{\delta}
\global\long\def\gD{\Delta}
\global\long\def\vph{\varphi}
\global\long\def\gf{\varphi}
\global\long\def\gl{\lambda}
\global\long\def\gk{\kappa}

\global\long\def\wh#1{\widehat{#1}}
\global\long\def\bv#1{\mathbf{#1}}
\global\long\def\bs#1{\boldsymbol{#1}}

\global\long\def\ui{\wh{\boldsymbol{\imath}}}
\global\long\def\uj{\wh{\boldsymbol{\jmath}}}
\global\long\def\uk{\wh{\textbf{\em k}}}

\global\long\def\bosy#1{\boldsymbol{#1}}

\global\long\def\vect#1{\overline{\mathbf{#1}}}

\global\long\def\uI{\widehat{\mathbf{I}}}
\global\long\def\uJ{\widehat{\mathbf{J}}}
\global\long\def\uK{\widehat{\mathbf{K}}}

\global\long\def\uv#1{\widehat{\mathbf{#1}}}

\global\long\def\cross{\boldsymbol{\times}}

\global\long\def\ddt#1{\frac{\dee#1}{\dee t}}
\global\long\def\dbyd#1#2{\frac{\dee#2}{\dee#1}}
\global\long\def\parby#1#2{\frac{\partial#1}{\partial#2}}

\newcommandx\parder[2][usedefault, addprefix=\global, 1=]{\frac{\partial#2}{\partial#1}}

\global\long\def\fall{,\quad\text{for all}\quad}

\global\long\def\reals{\mathbb{R}}

\global\long\def\rthree{\reals^{3}}
\global\long\def\rsix{\reals^{6}}
\global\long\def\rn{\reals^{n}}
\global\long\def\prn{\reals^{n+}}
\global\long\def\nrn{\reals^{n-}}
\global\long\def\cprn{\overline{\reals}^{n+}}
\global\long\def\cnrn{\overline{\reals}^{n-}}

\global\long\def\les{\leqslant}

\global\long\def\ges{\geqslant}

\global\long\def\dee{\mathrm{d}}

\global\long\def\from{\colon}

\global\long\def\tto{\longrightarrow}

\global\long\def\abs#1{\left|#1\right|}

\global\long\def\isom{\cong}

\global\long\def\comp{\circ}

\global\long\def\cl#1{\overline{#1}}

\global\long\def\fun{\varphi}

\global\long\def\interior{\mathrm{Int\,}}
\global\long\def\diver{\mathrm{div\,}}

\global\long\def\sign{\mathrm{sign\,}}
\global\long\def\sgn#1{(-1)^{#1}}
\global\long\def\sgnp#1{(-1){}^{\abs{#1}}}

\global\long\def\dimension{\mathrm{dim\,}}

\global\long\def\esssup{\mathrm{ess}\,\sup}

\global\long\def\ess{\mathrm{{ess}}}

\global\long\def\kernel{\text{Kernel}\,}

\global\long\def\support{\mathrm{Supp}\,}

\global\long\def\image{\mathrm{Image\,}}
\global\long\def\oper#1{\mathrm{#1}\,}

\global\long\def\emb#1{\mathrm{Emb}(#1)}

\global\long\def\resto#1#2{\left.#1\right|_{#2}}

\global\long\def\incl{\iota}

\global\long\def\rest{\rho}
\global\long\def\extnd{e_{0}}

\global\long\def\proj{\pi}

\global\long\def\sphere{S^{2}}
\global\long\def\hemis{H}

\global\long\def\ino#1{\int\limits _{#1}}

\global\long\def\half{\frac{1}{2}}

\global\long\def\shalf{{\scriptstyle \half}}

\global\long\def\third{\frac{1}{3}}

\global\long\def\empt{\varnothing}

\global\long\def\paren#1{\left(#1\right)}

\global\long\def\bigp#1{\bigl(#1\bigr)}

\global\long\def\biggp#1{\biggl(#1\biggr)}

\global\long\def\Bigp#1{\Bigl(#1\Bigr)}

\global\long\def\braces#1{\left\{  #1\right\}  }

\global\long\def\sqbr#1{\left[#1\right]}

\global\long\def\norm#1{\|#1\|}

\global\long\def\trps{^{\mathsf{T}}}

\global\long\def\alt{\mathfrak{A}}

\global\long\def\pou{\eta}

\global\long\def\ext{\bigwedge}
\global\long\def\forms{\Omega}

\global\long\def\dotwedge{\dot{\mbox{\ensuremath{\wedge}}}}

\global\long\def\vel{\theta}

\global\long\def\contr{\raisebox{0.4pt}{\mbox{\ensuremath{\lrcorner}}}\,}

\global\long\def\fcontr{\raisebox{0.4pt}{\mbox{\ensuremath{\llcorner}}}\,}

\global\long\def\lie{\mathcal{L}}

\global\long\def\jet#1{j^{1}(#1)}

\global\long\def\Jet#1{J^{1}(#1)}

\global\long\def\L#1{L\bigl(#1\bigr)}

\global\long\def\vvforms{\ext^{\dims}\bigp{T\spc,\vbts^{*}}}

\global\long\def\contr{\raisebox{0.4pt}{\mbox{\ensuremath{\lrcorner}}}\,}

\global\long\def\lisub#1#2#3{#1_{1}#2\dots#2#1_{#3}}

\global\long\def\lisup#1#2#3{#1^{1}#2\dots#2#1^{#3}}

\global\long\def\lisubb#1#2#3#4{#1_{#2}#3\dots#3#1_{#4}}

\global\long\def\lisubbc#1#2#3#4{#1_{#2}#3\cdots#3#1_{#4}}

\global\long\def\lisubbwout#1#2#3#4#5{#1_{#2}#3\dots#3\widehat{#1}_{#5}#3\dots#3#1_{#4}}

\global\long\def\lisubc#1#2#3{#1_{1}#2\cdots#2#1_{#3}}

\global\long\def\lisupc#1#2#3{#1^{1}#2\cdots#2#1^{#3}}

\global\long\def\lisupp#1#2#3#4{#1^{#2}#3\dots#3#1^{#4}}

\global\long\def\lisuppc#1#2#3#4{#1^{#2}#3\cdots#3#1^{#4}}

\global\long\def\lisuppwout#1#2#3#4#5#6{#1^{#2}#3#4#3\wh{#1^{#6}}#3#4#3#1^{#5}}

\global\long\def\lisubbwout#1#2#3#4#5#6{#1_{#2}#3#4#3\wh{#1}_{#6}#3#4#3#1_{#5}}

\global\long\def\lisubwout#1#2#3#4{#1_{1}#2\dots#2\widehat{#1}_{#4}#2\dots#2#1_{#3}}

\global\long\def\lisupwout#1#2#3#4{#1^{1}#2\dots#2\widehat{#1^{#4}}#2\dots#2#1^{#3}}

\global\long\def\lisubwoutc#1#2#3#4{#1_{1}#2\cdots#2\widehat{#1}_{#4}#2\cdots#2#1_{#3}}

\global\long\def\twp#1#2#3{\dee#1^{#2}\wedge\dee#1^{#3}}

\global\long\def\thp#1#2#3#4{\dee#1^{#2}\wedge\dee#1^{#3}\wedge\dee#1^{#4}}

\global\long\def\fop#1#2#3#4#5{\dee#1^{#2}\wedge\dee#1^{#3}\wedge\dee#1^{#4}\wedge\dee#1^{#5}}

\global\long\def\idots#1{#1\dots#1}
\global\long\def\icdots#1{#1\cdots#1}

\global\long\def\pis{x}

\global\long\def\pib{X}

\global\long\def\body{\mathcal{B}}

\global\long\def\man{\mathcal{M}}

\global\long\def\srfc{\mathcal{S}}

\global\long\def\hyper{\mathcal{S}}

\global\long\def\bdry{\partial}

\global\long\def\gO{\varOmega}

\global\long\def\reg{R}

\global\long\def\bdom{\bdry\reg}

\global\long\def\bndo{\partial\gO}
\global\long\def\pbndo{\Gamma}
\global\long\def\bndoo{\pbndo_{0}}
 \global\long\def\bndot{\pbndo_{t}}

\global\long\def\cloo{\cl{\gO}}

\global\long\def\nor{\boldsymbol{n}}
\global\long\def\nora{\nor}
\global\long\def\norb{\boldsymbol{u}}
\global\long\def\norc{v}

\global\long\def\dA{\,\dee A}

\global\long\def\dV{\,\dee V}

\global\long\def\eps{\varepsilon}

\global\long\def\vs{\mathbf{W}}
\global\long\def\avs{\mathbf{V}}

\global\long\def\vbase{\boldsymbol{e}}
\global\long\def\sbase{\mathbf{e}}
\global\long\def\mvec{\mathfrak{v}}

\global\long\def\inter{\mathcal{I}}

\global\long\def\motn{m}

\global\long\def\conf{\kappa}

\global\long\def\vf{w}
\global\long\def\flow{\Phi}

\global\long\def\avf{u}

\global\long\def\stn{\varepsilon}

\global\long\def\rig{r}

\global\long\def\rigs{\mathcal{R}}

\global\long\def\qrigs{\!/\!\rigs}

\global\long\def\qd{\!/\,\!\kernel\diffop}

\global\long\def\dis{\chi}

\global\long\def\fc{F}

\global\long\def\st{\sigma}

\global\long\def\bfc{b}

\global\long\def\sfc{t}

\global\long\def\stm{S}

\global\long\def\sts{\varSigma}

\global\long\def\ebdfc{T}
\global\long\def\optimum{\st^{\mathrm{opt}}}
\global\long\def\scf{K}

\global\long\def\curr{T}

\global\long\def\forms{\Omega}

\global\long\def\cee#1{C^{#1}}

\global\long\def\lone{L^{1}}

\global\long\def\linf{L^{\infty}}

\global\long\def\lp#1{L^{#1}}

\global\long\def\ofbdo{(\bndo)}

\global\long\def\ofclo{(\cloo)}

\global\long\def\vono{(\gO,\rthree)}

\global\long\def\vonbdo{(\bndo,\rthree)}
\global\long\def\vonbdoo{(\bndoo,\rthree)}
\global\long\def\vonbdot{(\bndot,\rthree)}

\global\long\def\vonclo{(\cl{\gO},\rthree)}

\global\long\def\strono{(\gO,\reals^{6})}

\global\long\def\sob{W_{1}^{1}}

\global\long\def\sobb{\sob(\gO,\rthree)}

\global\long\def\lob{\lone(\gO,\rthree)}

\global\long\def\lib{\linf(\gO,\reals^{12})}

\global\long\def\ofO{(\gO)}

\global\long\def\oneo{{1,\gO}}
\global\long\def\onebdo{{1,\bndo}}
\global\long\def\info{{\infty,\gO}}

\global\long\def\infclo{{\infty,\cloo}}

\global\long\def\infbdo{{\infty,\bndo}}

\global\long\def\ld{LD}

\global\long\def\ldo{\ld\ofO}
\global\long\def\ldoo{\ldo_{0}}

\global\long\def\trace{\gamma}

\global\long\def\pr{\proj_{\rigs}}

\global\long\def\pq{\proj}

\global\long\def\qr{\,/\,\reals}

\global\long\def\aro{S_{1}}
\global\long\def\art{S_{2}}

\global\long\def\mo{m_{1}}
\global\long\def\mt{m_{2}}

\global\long\def\yieldc{B}

\global\long\def\yieldf{Y}

\global\long\def\trpr{\pi_{P}}

\global\long\def\devpr{\pi_{\devsp}}

\global\long\def\prsp{P}

\global\long\def\devsp{D}

\global\long\def\ynorm#1{\|#1\|_{\yieldf}}

\global\long\def\colls{\Psi}

\global\long\def\ssx{S}

\global\long\def\smap{s}

\global\long\def\smat{\chi}

\global\long\def\sx{e}

\global\long\def\snode{P}

\global\long\def\elem{e}

\global\long\def\nel{L}

\global\long\def\el{l}

\global\long\def\ipln{\phi}

\global\long\def\ndof{D}

\global\long\def\dof{d}

\global\long\def\nldof{N}

\global\long\def\ldof{n}

\global\long\def\lvf{\chi}

\global\long\def\lfc{\varphi}

\global\long\def\amat{A}

\global\long\def\snomat{E}

\global\long\def\femat{E}

\global\long\def\tmat{T}

\global\long\def\fvec{f}

\global\long\def\snsp{\mathcal{S}}

\global\long\def\slnsp{\Phi}

\global\long\def\ro{r_{1}}

\global\long\def\rtwo{r_{2}}

\global\long\def\rth{r_{3}}

\global\long\def\mind{\alpha}
\global\long\def\vb{\xi}

\global\long\def\vbt{E}
\global\long\def\fib{\mathbf{V}}

\global\long\def\jetb#1{J^{#1}}

\global\long\def\jetm#1{j_{#1}}

\global\long\def\sobp#1#2{W_{#2}^{#1}}

\global\long\def\inner#1#2{\left\langle #1,#2\right\rangle }

\global\long\def\fields{\sobp pk(\vb)}

\global\long\def\bodyfields{\sobp p{k_{\partial}}(\vb)}

\global\long\def\forces{\sobp pk(\vb)^{*}}

\global\long\def\bfields{\sobp p{k_{\partial}}(\vb\resto{\bndo})}

\global\long\def\loadp{(\sfc,\bfc)}

\global\long\def\strains{\lp p(\jetb k(\vb))}

\global\long\def\stresses{\lp{p'}(\jetb k(\vb)^{*})}

\global\long\def\diffop{D}

\global\long\def\strainm{E}

\global\long\def\incomps{\vs_{\yieldf}}

\global\long\def\devs{L^{p'}(\eta_{1}^{*})}

\global\long\def\incompsns{L^{p}(\eta_{1})}

\global\long\def\dists{\mathcal{D}'}
\global\long\def\testfs{\mathcal{D}}

\global\long\def\prop{P}

\global\long\def\aprop{Q}

\global\long\def\flux{T}

\global\long\def\fform{\tau}

\global\long\def\dimn{n}

\global\long\def\sdim{{\dimn-1}}

\global\long\def\prodf{{\scriptstyle \smallsmile}}

\global\long\def\ptnl{\varphi}

\global\long\def\form{\omega}

\global\long\def\dens{\rho}

\global\long\def\simp{s}

\global\long\def\cell{C}

\global\long\def\chain{B}

\global\long\def\ach{A}

\global\long\def\coch{X}

\global\long\def\scale{s}

\global\long\def\fnorm#1{\abs{#1}^{\flat}}

\global\long\def\chains{\mathcal{A}}

\global\long\def\ivs{\boldsymbol{U}}

\global\long\def\mvs{\boldsymbol{V}}

\global\long\def\cvs{\boldsymbol{W}}

\global\long\def\subbs{\mathcal{B}}

\global\long\def\elements{\mathcal{E}}

\global\long\def\element{E}

\global\long\def\nodes{\mathcal{N}}

\global\long\def\node{N}

\global\long\def\psubbs{\mathcal{P}}

\global\long\def\psubb{P}

\global\long\def\matr{M}

\global\long\def\nodemap{\nu}

\global\long\def\dform{\psi}

\global\long\def\R{\reals}
\global\long\def\Z{\mathbb{Z}}

\title[Geometry and Kinematics of Singular Defects]{On the Geometry and Kinematics of Smoothly Distributed and Singular
Defects}

\author{Marcelo Epstein and Reuven Segev}

\address{University of Calgary, Canada; Ben-Gurion University of the Negev,
Israel }

\email{mepstein@ucalgary.ca; rsegev@bgu.ac.il}


\begin{abstract}
A continuum mechanical framework for the description of the geometry
and kinematics of defects in material structure is proposed. The setting
applies to a body manifold of any dimension which is devoid of a Riemannian
or a parallelism structure. In addition, both continuous distributions
of defects as well as singular distributions are encompassed by the
theory. In the general case, the material structure is specified by
a de Rham current $\curr$ and the associated defects are given by
its boundary $\bdry\curr$. For a motion of defects associated with
a family of diffeomorphisms of a material body, it is shown that the
rate of change of the distribution of defects is given by the dual
of the Lie derivative operator.
\end{abstract}
\maketitle

\section{Introduction}

We present below a mathematical framework for the description of the
geometry and kinematics of material defects from the continuum mechanics,
macroscopic, point of view. In particular, the proposed framework
applies to both continuously distributed as well as singular defects
and is formulated on general manifolds devoid of any metric or a parallelism
structure. 

Material defects, are frequently described by relative deformation
of neighboring points in the material (\eg, \cite{kronerAnthony75,LurieAndKalamkarov2006,Sahoo84}).
 Sometimes a global point of view is adopted (\eg, \cite{Cermelli99})
and defects are viewed as obstructions to the construction of a global
inverse deformation. Another frequent approach (\eg, \cite{kondo,noll,wang,epselz}),
views the existence of defects, or inhomogeneities, as an inherent
consequence of the constitutive relation for a body. Following \cite{ESdislocations2012},
the present framework differs from the first point of view above in
the sense that the analysis involves no kinematics of the body in
space. No deformations are considered and only the material structure
of the body manifold is studied. The present approach differs from
the theory of inhomogeneities in the sense that rather than associating
the defects with a particular constitutive relation, \eg, the relation
between the stress and the deformation, the material structure is
given explicitly. (See somewhat similar approaches in by Toupin \cite{Toupin1968}
and Eringen \& Claus \cite{EringenAndClaus1970} who use oriented,
or micromorphic media.) For example, it is assumed that a family of
Bravais hyperplanes is given by explicitly prescribing, at each point
in a body, a hyperplane as well as the density of these hyperplanes.
In other words, one specifies a distributed analog of the Miller indices
for a family of hyperplanes. Mathematically speaking, if the body
is a manifold $\man$ of dimension $n$, we consider in the continuous
case a distribution, an $(n-1)$-subbundle of the tangent bundle,
which is induced by a differential $1$-form $\gf$. Material structure
of dimension $r\ne n-1$, will be prescribed by a $p$-form for $p=n-r$.
Singular material structure of dimension $r$ is given in terms of
a de Rham $r$-current $\curr$, a generalized $(n-r)$-form. Thus,
for a $3$-dimensional manifold, the interesting cases are $r=2$
that gives the Bravais planes at the various points and $r=1$ that
gives the inclination field of directors for the theory of disclinations.

In the deformation theory of dislocations, the Burgers vector is defined
using the gap that opens up between the positions of neighboring points.
Here, one considers the total amount of hyperplanes that are penetrated,
in one particular orientation, when a closed loop is followed. This
motivates the definition of the distribution of defects as the exterior
derivative $\dee\gf$ of the structure form $\gf$ in the continuous
case and as the boundary $\bdry\curr$ of the structure current $\curr$
in the singular case.

An attempt was made here to introduce some of the relevant background
on differential forms and de Rham currents. In Section \ref{sec:FormsAndHypersurfaces}
we briefly review the subject of distributions, subbundles of the
tangent bundle, induced by a decomposable differential forms and the
results pertaining to the submanifolds they may induce. In Section
\ref{sec:Structure-FormsAndDefects} we use these results to introduce
structure $p$-forms and the corresponding exterior derivatives that
represent the associated smoothly distributed defects. Section \ref{sec:De-Rham-Currents}
presents the basic notions concerning de Rham currents and Section
\ref{sec:StructureAndDefectCurrents} uses these notions to introduce
the singular counterpart of structure forms and continuously distributed
defects. The simple cases of $0$-dimensional material structures
and $n$-dimensional material structures are considered in Section
\ref{sec:The-Simple-Cases}. Sections \ref{sec:Dislocations} and
\ref{sec:Disclinations} present the physically relevant cases of
dislocations and disclinations, and some examples are given. Thus,
all the cases relevant to the $3$-dimensional space are covered.
Section \ref{sec:Kinematics} considers the motion of material structure
and the associated defects, and the rate of change of the motion.
Both the continuous and singular cases are discussed for the case
where the material structure is carried with a family of diffeomorphism
of the body manifold. Finally, we give an example in which a smooth
distribution of defects evolves into a singular defect.

\section{Forms and Hypersurfaces\label{sec:FormsAndHypersurfaces}}

Defects are considered in this article to be obstacles to integrability.
Consider a family $\mathcal{F}$ of $(n-r)$-dimensional oriented
hypersurfaces in the body manifold $\man$. We view the family $\mathcal{F}$
as a given material structure in the body. For example, a family of
$2$-dimensional surfaces in a $3$-dimensional body may be thought
of as a family of lattice layers. Let $\hyper$ be an $(r+1)$-dimensional
submanifold with boundary in the body manifold. The ``amount'' hypersurfaces
belonging to $\mathcal{F}$ that cross the boundary $\bdry\hyper$,
if different from zero, indicates the generation or annihilation of
such hypersurfaces in $\hyper$. We view such creation or annihilation
of material hypersurfaces as an indication for the presence of defects
in $\hyper$.

In this section, we describe the notions from exterior calculus used
for the description of what is referred to above as a ``family of
hypersurfaces'' in the body manifold $\man$.

We recall \cite[pp.~16--17]{Sternberg} that an $r$-dimensional subspace
$W$ of a vector space $V$ is associated with a decomposable $r$-vector
$\mvec$ which is unique up to a scalar factor such that $u\in W$
if and only if $\mvec\wedge u=0$. In the sequel we will use this
property for subspaces $D_{x}^{*}$ of the various cotangent spaces
$T_{x}^{*}\man$, $x\in\man$ of dimension $p=n-r$. It is observed
that each $D_{x}^{*}$ determines a unique $r$-dimensional subspace
$D_{x}=(D_{x}^{*})^{\perp}=\{v\in T_{x}\man\mid\go(v)=0,\,\text{for all }\go\in D_{x}^{*}\}$.
Thus, a $p$-dimensional subspace $D_{x}^{*}\subset T_{x}^{*}\man$
is determined by a decomposable $p$-covector (alternating tensor)
$\vph$. 

We will use the notation $v\contr\form$ for the contraction of a
$p$-covector $\form$ with a vector $v$, a $(p-1)$-covector satisfying
\begin{equation}
v\contr\form(\vf_{1},\dots,\vf_{p-1})=\form(v,\vf_{1},\dots,\vf_{p-1})=\form(v\wedge\lisubbc{\vf}1{\wedge}{p-1}).\label{eq:ContractionDefined}
\end{equation}
Let $\go$ be a $q$-covector, $q<p$, such that $\vph\wedge\go=0$.
Then, recalling the identity 
\begin{equation}
v\contr(\vph\wedge\go)=(v\contr\vph)\wedge\go+\sgn r\vph\wedge(v\contr\go),
\end{equation}
for any tangent vector $v$, one has, 
\begin{equation}
\vph\wedge(v\contr\go)=0
\end{equation}
for every vector $v$ that annihilates $\vph$ in the sense that $v\contr\vph=0$.
In the particular case $q=1$, and assuming $\vph\ne0$, one has $\go(v)=0$. 

Conversely, if $\go(v)=0$ for every $v\in D_{x}$, 
\begin{equation}
(v\contr\vph)\wedge\go=0
\end{equation}
for every covector $\go$ such that $\vph\wedge\go=0$. Since $\vph$
is decomposable, there is a basis $\{\gf^{1},\dots,\gf^{n}\}$ of
$T_{x}\man$ such that $\gf$ can be expressed as $\vph=\lisupc{\vph}{\wedge}p$.
Hence,
\begin{equation}
v\contr\vph=\sum_{i=1}^{p}\sgn{i-1}v^{i}\lisuppwout{\vph}1{\wedge}{\cdots}pi.
\end{equation}
For any $j=1,\dots,p$, $\gf\wedge\gf^{j}=0$, hence, 
\begin{equation}
0=(v\contr\vph)\wedge\vph^{j}=\sgn{p-j-1}v^{j}\lisupc{\vph}{\wedge}p.
\end{equation}
It follows that $v^{j}=0$, for all $j=1,\dots,p$, so that $v\contr\vph=0$.
We conclude that $v\in D_{x}$ if and only if 
\begin{equation}
v\contr\vph=0\label{eq:OrthogonalHyperplaneByContr}
\end{equation}
 and $\dimension D_{x}=r=n-p$.

\begin{rem}
Clearly, in a dual procedure and as given in \cite[pp.~16--17]{Sternberg},
one could start with a simple $r$-vector $\mvec$ at a point $x\in\man$
and define the subspace $D_{x}=\{v\in T_{x}\man\mid\mvec\wedge v=0\}$.
Then, the orthogonal subspace is given by $D_{x}^{*}=\{\ga\in T_{x}^{*}\man\mid\ga(v)=0,\; v\in D_{x}\}$.
For an $r$-vector $\mvec$ and a $k$-covector $\form$, with $r\ges k$,
we use the inner product notation $\mvec\fcontr\form$, an $(r-k)$-vector
defined by,
\begin{equation}
\gf(\mvec\fcontr\form)=(\gf\wedge\form)(\mvec),
\end{equation}
for every $(r-k)$-covector $\gf$. The condition that $\ga\in D_{x}^{*}$
may then be written as 
\begin{equation}
\mvec\fcontr\ga=0.\label{eq:DualOrthoByContraction}
\end{equation}

\end{rem}

A smooth decomposable differential $p$-form $\gf$ will induce therefore
a distribution $D$ on $\man$ of dimension $r=n-p$. Here, by a ``distribution''
we mean a subbundle of the tangent bundle rather than a Schwartz distribution.
Conversely, a distribution $D$ of dimension $r=n-p$ will induce
a collection of forms such that if $\gf$ induces $D$, so would the
form $a\gf$ for any positive, real valued function $a$ on $\man$.

Let $v_{1},\dots,v_{p}\in T_{x}\man$. We interpret $\gf(x)(v_{1},\dots,v_{p})$
as the amount of hyperplanes belonging to the distribution that cross
the infinitesimal $p$-dimensional oriented element (a $p$-dimensional
parallelepiped or a simplex) generated by the vectors $v_{1},\dots,v_{p}$.
In particular, if for some $i=1,\dots,p$, $v_{i}\in D_{x}$, so that
$v\contr\gf(x)=0$, this quantity will vanish as the hyperplanes and
the subspace generated by $v_{1},\dots,v_{p}$ intersect on a subspace
of dimension greater than zero. Multiplying the form $\gf$ by a positive
function $a$, the resulting form $a\gf$ is interpreted as describing
a family of hyperplanes which are parallel to those represented by
$\gf$, and whose density $a$ times larger.

A distribution does not represent necessarily tangent spaces to a
family of hypersurfaces, as we wish to consider. It is recalled that
an $r$-dimensional submanifold $\hyper$ is an integral manifold
of the distribution if $T_{x}\hyper=D_{x}$ for all $x\in\hyper$.
A distribution $D$ is referred to as involutive if at each $x\in\man$,
$D_{x}$ is the tangent space of an $r$-dimensional integral manifold.
The Frobenius theorem implies (\eg, \cite[pp.~441-442]{AbeMarsdenManifolds})
that the distribution $D$ is involutive if and only if there is a
$1$-form $\gb$ on $\man$ such that 
\begin{equation}
\dee\gf=\gb\wedge\gf.\label{eq:IntegrabilityCondition}
\end{equation}

Consider the form $\gf_{a}=a\gf$ for a function $a$. Recalling the
identity 
\begin{equation}
\dee(\mu\wedge\nu)=\dee\mu\wedge\nu+(-1)^{q}\mu\wedge\dee\nu,\label{eq:ExterDerOfExterProd}
\end{equation}
for the $q$-form $\mu$ and a form $\nu$ over $\man$, one has
\begin{equation}
\dee\gf_{a}=\dee a\wedge\gf+a\dee\gf.\label{eq:derFactoredForm}
\end{equation}
Assume that condition (\ref{eq:IntegrabilityCondition}) holds. Then,
if the function $a$ is a solution of the equation $\dee a=-a\gb$,
(\ref{eq:derFactoredForm}) implies that $\dee\gf_{a}=0$. Conversely,
assume that $\dee\gf_{a}=0$ for some positive function $a$. Then,
the one form $\gb=-\dee a/a$ satisfies the integrability condition
(\ref{eq:IntegrabilityCondition}). We conclude therefore that the
distribution induced by a form $\gf$ is involutive if and only if
it has an integrating factor, a function $a$ on $\man$ such that
$\dee(a\gf)=0$. Thus, for a form that induces an involutive distribution,
the density of the hyperplanes at each point may be readjusted so
that the exterior derivative of the resulting form vanishes. In particular,
if $\dee\gf=0$, the distribution induced by $\gf$ is involutive.
\begin{rem}
\label{rem:NoEquivalentToExteriorDerForMultVecs}Let $D$ be a distribution
induced by a simple $r$-vector field $\mvec$ represented locally
by $v_{1}\wedge\cdots\wedge v_{r}$ for smooth vector fields $v_{1},\dots,v_{r}$.
Then, using the Lie bracket notation, the condition that the distribution
is involutive is that $[v_{i},v_{j}]$ is also a section of $D$ for
all $i,j=1,\dots,r$. It is noted however that we did not write a
condition yet on $\mvec$ that will be equivalent to the condition
$\dee\gf=0$. The theory of de Rham currents provides one with the
required tools for writing such a condition.
\end{rem}

\section{Structure Forms, Defect Forms and the Corresponding Frank's Rule\label{sec:Structure-FormsAndDefects}}

From the point of view of the material structure of bodies, any decomposable
$p$-form represents a distribution of hyperplanes, Bravais hyperplanes,
at the various points in the body while an involutive distribution
represents a collection of submanifolds at the various material points
\ie,  hyperplanes at various points may be assembled to form tangent
spaces of $n-p=r$-dimensional submanifolds---the material or Bravais
hypersurfaces. We will refer to such decomposable forms as \emph{structure
forms}. 

The material structure described by an involutive structure $p$-form
may still contain defects. Such defects are due to the creation or
loss of material hypersurfaces in some regions in the body. Let $\hyper$
be an $(r+1)$-dimensional manifold with a boundary. The creation
or loss of material hypersurfaces inside $\hyper$ will be reflected
by the integral of the structure form over the boundary, $\bdry\hyper$.
Note that the integrals of a form over $\hyper$ and its boundary
make sense even if the form is not involutive. In this case, the integral
over the boundary may naturally be interpreted as the creation of
hyperplanes rather than hypersurfaces.

Stokes's theorem asserts that 
\begin{equation}
\int_{\bdry\hyper}\gf=\int_{\hyper}\dee\gf.\label{eq:StokesThm}
\end{equation}
Thus, if the exterior derivative $\dee\gf$ of the structure form
vanishes, the total creation or annihilation of material hypersurfaces
within any $(r+1)$-submanifold $\hyper$, as reflected in the total
amount of hypersurfaces that cross the boundary $\bdry\hyper$, will
vanish. In other words, for a decomposable, form $\gf$ satisfying
(\ref{eq:IntegrabilityCondition}), which, by the Frobenius theorem,
induces a family of hypersurfaces, the stronger condition, $\dee\gf=0$,
\ie, $\gf$ is closed, implies that the family of hypersurfaces have
no sources or sinks. This suggests that $\dee\gf$ is the measure
of the sources of material, or Bravais, hypersurfaces inside the body
$\man$---the measure of the distribution of defects. We will refer
to $\dee\gf$ as the \emph{defect form} corresponding to $\gf$.

It is recalled that the skew symmetry of the exterior derivative combined
with the symmetry of second derivatives of functions implies that
for any form $\ga$, 
\begin{equation}
\dee^{2}\ga:=\dee(\dee\ga)=0.
\end{equation}
Let $\dform=\dee\gf$ be the defect form associated with the structure
form $\gf$. It follows, therefore, that $\dform$ must satisfy the
condition
\begin{equation}
\dee\dform=0.
\end{equation}
This compatibility condition is the analog of Frank's rules for defects
of any dimension on manifolds, as long they are smoothly distributed.

\section{De Rham Currents\label{sec:De-Rham-Currents}}

Let $\gf$ be a decomposable $p$-covector at a point $x\in\man$.
It follows that one may choose a basis $\{\vbase_{i}\}$, $i=1,\dots,n$,
of $T_{x}\man$ with dual basis $\{\gf^{i}\}$ such that $\gf=\lisupc{\gf}{\wedge}p$.
Let $\form$ be an $(n-p)$-covector such that $\gf\wedge\form\ne0$.
Then, $\form$ must be of the form $\form=a\lisuppc{\gf}{p+1}{\wedge}n+\ga$,
with $\gf\wedge\ga=0$, for some nonvanishing number $a$. The subspaces
induced by $\gf$ are spanned by $\{\vbase_{p+1},\dots,\vbase_{n}\}$.
Let $\{v_{1},\dots,v_{n}\}$ be $n$ vectors in $T_{x}\man$ and consider
$\gf\wedge\form(v_{1},\dots,v_{n})=\gf\wedge\form(v_{1}\wedge\cdots\wedge v_{n})$.
Then, $\lisubc v{\wedge}n$ must be of the form
\begin{equation}
\lisubc v{\wedge}n=b\lisubc{\vbase}{\wedge}n,
\end{equation}
for some real number $b$. The orthogonality of the two forms with
vectors in the respective subspaces implies that 
\begin{equation}
\gf\wedge\form(v_{1}\wedge\cdots\wedge v_{n})=\gf(\vbase_{1},\dots,\vbase_{p})\form(b\lisubbc{\vbase}{p+1}{\wedge}n)=ab.
\end{equation}
This quantity, as well as the identical $((\lisubbc v1{\wedge}n)\fcontr\gf)(\form)$,
is interpreted as the amount of cells formed by the hyperplanes induced
by the forms $\gf$ and $\form$ contained in the $n$-parallelepiped
determined by $v_{1},\dots,v_{n}$. 

Accordingly, for a $p$-form $\gf$ and an $(n-p)$-form $\form$,
one may interpret the integral
\begin{equation}
\int_{\man}\gf\wedge\form,
\end{equation}
as the total amount of cells in $\man$.

Therefore, one may consider the linear operator $\curr_{\gf}$ acting
on $(n-p)$-forms by
\begin{equation}
\curr_{\gf}(\form)=\int_{\man}\gf\wedge\form.\label{eq:SmoothCurr}
\end{equation}
whose action on an $(n-p)$-form $\form$ gives the total amount of
cells corresponding to $\gf\wedge\form$ in $\man$. 

A linear functional $\curr_{\gf}$ acting on differential forms as
in (\ref{eq:SmoothCurr}) is a typical simple example of a de Rham
current.

A de Rham $r$-current is a linear operator acting on the space of
smooth $r$forms with compact supports. A de Rham current $\curr$
is required to be continuous in the following sense. Let $(\form_{k})$
is a sequence of $r$-forms whose supports are all contained in a
compact subset of a coordinate neighborhood and whose local representatives
as well as all the partial derivatives of all orders of the local
representatives tend to zero uniformly as $k\to\infty$. Then, $\curr(\form_{k})\to0$.
Thus, for the case $r=0$, $\curr$ is a Schwartz distribution on
the manifold $\man$. For $r>0$, currents contain additional geometric
properties in comparison with Schwartz distributions.

In contrast with the example above where the $r$-current $\curr_{\gf}$
was induced by a smooth $(n-r)$-form $\gf$, currents may exhibit
singular behavior. As a typical simple example, an $r$-dimensional
submanifold $\hyper\subset\man$ induces a current $\curr_{\hyper}$
defined by
\begin{equation}
\curr_{\hyper}(\form)=\int_{\hyper}\form\label{eq:CurrBySubmanifold}
\end{equation}
for every $r$-form $\form$ with compact support. In comparison with
(\ref{eq:SmoothCurr}), the current $\curr_{\hyper}$ may be viewed
as the limit of currents of the form $\curr_{\gf}$ where the support
of $\gf$ shrinks to a small neighborhood of $\hyper$ and the value
of its components tend to infinity in that neighborhood. This process
may be made rigorous by the process of regularization (\eg,  \cite[pp.~61--70]{derham})
which is a generalization of the analogous process for Schwartz distributions.

A current may be restricted to the domain of a chart on $\man$ by
restricting its action to forms supported in the domain of that chart.
An $r$-form $\form$ which is supported in the domain of a chart,
may be expressed using real valued functions $\form_{\mu}$ as 
\begin{equation}
\form=\sum_{(\mu)}\form_{\mu}\dee x^{\mu},
\end{equation}
where $\mu$ is an increasing (indicated by the parenthesis around
it) $r$-multi-index taking values in the range $1,\dots,n$. By linearity,
\begin{equation}
\curr(\form)=\sum_{(\mu)}\curr(\form_{\mu}\dee x^{\mu})=\sum_{(\mu)}\curr^{\mu}(\form_{\mu}),
\end{equation}
where $\curr^{\mu}$ are the Schwartz distributions, $0$-currents,
so that $\curr^{\mu}(\form_{\mu})=\curr(\form_{\mu}\dee x^{\mu})$.
For an $m$-vector field $\mvec$ and an $r$-current $\curr$, consider
the $(r+m)$-current $\curr\wedge\mvec$ defined by 
\begin{equation}
(\curr\wedge\mvec)(\form)=\curr(\mvec\contr\form).\label{eq:currentContrMultiVec}
\end{equation}
Here, in analogy with (\ref{eq:ContractionDefined}) $\mvec\contr\form$
is the $r$-form such that for any $r$-vector field $\mathfrak{w}$,
\begin{equation}
\mvec\contr\form(\mathfrak{w})=\form(\mvec\wedge\mathfrak{w}).
\end{equation}
Then, the restriction of a current $\curr$ to a chart with domain
$U\subset\man$ may be represented locally by distributions $\curr^{\gl}$
in the form 
\begin{equation}
\resto{\curr}U=\sum_{(\gl)}\curr^{\gl}\wedge\parder[x^{\gl}]{}.\label{eq:CurrAsGeneralizedMultivector}
\end{equation}
This representation views a current as a generalized multivector field
which is the approach of \cite[p.~199]{Whitney1957}. Using a partition
of unity, a current may be represented by its restrictions to the
domains of charts.

For a smooth $m$-form $\ga$ and an $r$-current $\curr$ with $r>m$,
the $(r-m)$-current $\curr\fcontr\ga$ is defined by 
\begin{equation}
\curr\fcontr\ga(\form)=\curr(\ga\wedge\form).
\end{equation}
Using this notation, de Rham's representation of currents may be expressed
as follows. Let $\{\curr_{\mu}\}$, where $\mu$ is an increasing
multi-index with $\mu_{i}=1,\dots,n$, $i=1,\dots,n-r$, be $n$-currents
in a coordinate neighborhood. Consider the $r$-current
\begin{equation}
\curr=\sum_{(\mu)}\curr_{\mu}\fcontr\dee x^{\mu},\label{eq:deRhamRepres}
\end{equation}
so that 
\begin{equation}
\curr(\go)=\sum_{(\mu)}\curr_{\mu}(\dee x^{\mu}\wedge\go).\label{eq:deRhamRepres-b}
\end{equation}
Using the linearity of the currents, it may be shown that the restriction
of a current to a coordinate neighborhood may be represented in the
form (\ref{eq:deRhamRepres}). It is noted that in (\ref{eq:deRhamRepres}),
currents are viewed as generalized forms.

Let $\curr$ be a current which is given in a coordinate neighborhood
in $\man$ by the single Schwartz distribution $T^{0}$ and an $r$-vector
field $\mvec$ in the form 
\begin{equation}
T=T^{0}\wedge\mvec.\label{eq:DecomposableCurrent}
\end{equation}
Then, $\curr\fcontr\ga=0$ for every $1$-form $\ga$ that takes values
in the distribution $D^{*}$ which is orthogonal to that induced by
$\mvec$. Conversely, let $D^{*}$ be a $p$-dimensional subbundle
of $T^{*}\man$ and assume that for a current $\curr$, $\curr\fcontr\ga=0$
for every $1$-form $\ga$ valued in $D^{*}$. Then, $\curr$ is of
the form (\ref{eq:DecomposableCurrent}) where $\mvec$ is a multivector
that induces the distribution $D$ which is orthogonal to $D^{*}$.

The boundary of an $r$-current $\curr$ is the $(r-1)$-current $\bdry\curr$
defined by the condition
\begin{equation}
\bdry\curr(\form)=\curr(\dee\form).
\end{equation}
Consider the $r$-current $\curr_{\gf}$ defined in terms of a smooth
$(n-r)$-form $\gf$ as in (\ref{eq:SmoothCurr}). Then, using (\ref{eq:ExterDerOfExterProd}),
Stokes's theorem and the fact that $\form$ has a compact support
in $\man$, one has
\begin{equation}
\begin{split}\bdry\curr_{\gf}(\form) & =\int_{\man}\gf\wedge\dee\form,\\
 & =(-1)^{n-r}\left[\int_{\man}\dee(\gf\wedge\form)-\int_{\man}\dee\gf\wedge\form\right]\\
 & =(-1)^{n-r}\left[\int_{\bdry\man}\gf\wedge\form-\int_{\man}\dee\gf\wedge\form\right],\\
 & =(-1)^{n-r+1}\int_{\man}\dee\gf\wedge\form.
\end{split}
,
\end{equation}
It follows that, 
\begin{equation}
\bdry\curr_{\gf}=(-1)^{n-r+1}\curr_{\dee\gf}.\label{eq:BdryOfCurrIndcdBySmoothForm}
\end{equation}
For an $r$-dimensional submanifold with boundary $\hyper$, the boundary
of the current $\curr_{\hyper}$ defined in (\ref{eq:CurrBySubmanifold})
satisfies
\begin{equation}
\begin{split}\bdry\curr_{\hyper}(\form) & =\curr_{\hyper}(\dee\form),\\
 & =\int_{\hyper}\dee\form,\\
 & =\int_{\bdry\hyper}\form.
\end{split}
\end{equation}
Hence,
\begin{equation}
\bdry\curr_{\hyper}=\curr_{\bdry\hyper}\label{eq:BdryOfCurrIndcdBySubman}
\end{equation}
which motivates the terminology used.

Finally, since for every form $\form$, $\dee^{2}\form=0$, one has
$\bdry^{2}\curr(\form)=\bdry(\bdry\curr(\form))=\curr(\dee^{2}\form)$,
and we conclude that 
\begin{equation}
\bdry^{2}\curr=0,\label{eq:bdryNilpotent}
\end{equation}
identically.

\section{Structure Currents, Defect Currents and Frank's Rules\label{sec:StructureAndDefectCurrents}}

It is concluded from the previous section that de Rham currents may
be thought of as generalizations of smooth differential forms to the
singular, non-smooth, case, or alternatively, as generalization of
smooth multivector fields to the singular case. In addition, the boundary
of a current generalizes the exterior derivative of a form. Thus,
an $r$-current will be the singular counterpart of a $p=n-r$ structure
form and will be referred to as a structure current. Accordingly,
for a structure current $\curr$, the boundary $\bdry\curr$ will
represent the geometry of the defects and will be referred to as the
\emph{defect current}. The material structure represented by the current
$\curr$ will be defect free if $\bdry\curr=0$.

We recall that the constancy theorem for currents asserts that on
a connected manifold $\man$, a closed $n$-current $\curr$, \ie, 
$\curr$ satisfies $\bdry\curr=0$, is represented by a constant $c$
in the form
\begin{equation}
T(\form)=c\int_{\man}\form.
\end{equation}
One may apply this to the de Rham representation of currents (\ref{eq:deRhamRepres})
as follows. We observe first that by (\ref{eq:deRhamRepres}), one
has
\begin{equation}
\begin{split}\bdry\curr(\form) & =\curr(\dee\form)=\sum_{(\mu)}\curr_{\mu}(\dee x^{\mu}\wedge\dee\go),\\
 & =(-1)^{n-r}\sum_{(\mu)}\curr_{\mu}(\dee(\dee x^{\mu}\wedge\go)),\\
 & =(-1)^{n-r}\sum_{(\mu)}\bdry\curr_{\mu}(\dee x^{\mu}\wedge\go),\\
 & =(-1)^{n-r}\sum_{(\mu)}\bdry\curr_{\mu}\fcontr\dee x^{\mu}(\go),
\end{split}
\end{equation}
where (\ref{eq:ExterDerOfExterProd}) was used in the second line.
It follows that the boundary of the current $\curr$ can always be
represented by the $(n-1)$-currents $\bdry\curr_{\mu}$ in the form
\begin{equation}
\bdry\curr=(-1)^{n-r}\sum_{(\mu)}\bdry\curr_{\mu}\fcontr\dee x^{\mu}.\label{eq:DeRhamReprtnOfBdry}
\end{equation}

Assume that the current $\curr$ is given the form 
\begin{equation}
\curr=\curr_{0}\fcontr\gf=\curr_{0}\fcontr(\gf^{1}\wedge\cdots\wedge\gf^{n-r})
\end{equation}
for an $n$-current $\curr_{0}$ and a collection of $n-r$ linearly
independent$1$-forms $\gf^{i}$, $i=1,\dots,n-r$ spanning a subbundle
$D^{*}$of $T^{*}\man$. The current $\curr$ can thus be associated
with the distribution $D^{*}$. In particular, let $\psi$ be any
$1$-form valued in $D^{*}$, then, for each $(r-1)$-form $\go$,
\begin{equation}
\begin{split}(\curr\fcontr\psi)(\form) & =(\curr_{0}\fcontr\gf)(\psi\wedge\form),\\
 & =T_{0}(\gf\wedge\psi\wedge\form),\\
 & =0.
\end{split}
\end{equation}
Thus, $\curr\fcontr\psi=0$. 

Conversely, assume that for a general current $\curr$, we are given
that $\curr\fcontr\psi=0$ for every section $\psi$ of a subbundle
$D^{*}$. We consider the restriction of $\curr$ to a coordinate
neighborhood in which $D^{*}$ is induced by the form $\gf=\lisuppc{\gf}1{\wedge}{n-r}$
in which $\gf^{1},\dots,\gf^{n}$ span $T^{*}\man$. Writing $\curr=\sum_{(\gl)}T_{\gl}\fcontr\gf^{\gl}$,
it follows that for any $(r-1)$-form $\form$ and all sections $\psi$
of $D^{*}$,
\begin{equation}
\begin{split}0 & =(\curr\fcontr\psi)(\form),\\
 & =\sum_{(\gl)}((T_{\gl}\fcontr\gf^{\gl})\fcontr\psi)(\form),\\
 & =\sum_{(\gl)}\curr_{\gl}(\gf^{\gl}\wedge\psi\wedge\form).
\end{split}
\end{equation}
Since $\gf^{1\dots n-r}\wedge\psi=0$, it follows that $\curr_{\gl}=0$
for all $\gl\ne1,\dots,r$. Hence, $\curr=\curr_{1\dots n-r}\fcontr(\gf^{1}\wedge\cdots\wedge\gf^{n-r})$.
We conclude that $\curr\fcontr\psi=0$ for every section of a subbundle
$D^{*}\subset T^{*}\man$, if an only if 
\begin{equation}
\curr=\curr_{0}\fcontr\gf\label{eq:DecomposableCurrent-1}
\end{equation}
for an $n$-current $\curr_{0}$ and an $(n-r)$-form $\gf$ associated
with $D^{*}$. We will refer to such a current as a decomposable current.
It is observed that the condition $\curr\fcontr\psi=0$ for every
section $\psi$ of $D^{*}$, induces an ideal on the collection of
forms in the sense that for each $q$-form $\ga$, with $q\les n-r-1$,
$\curr\fcontr\psi=0$ for all $\psi$ implies that $\curr\fcontr(\psi\wedge\ga)=(\curr\fcontr\psi)\wedge\ga=0$
also.

Finally, if $S$ is a current representing the defect structure, the
identity $\bdry^{2}\curr=0$ implies that $\bdry S=0$, necessarily.
This is the generalization of Frank's rules for a possibly singular
defect structure.

\section{The Simple Cases\label{sec:The-Simple-Cases}}

In this section we consider the simple, possibly trivial, cases of
$n$-currents and $0$-currents, where it is recalled that $0$-currents
are Schwartz distributions on the manifold $\man$.

\subsection{$0$-Forms, $n$-Currents and Nonuniformity\label{sec:n-currents}}

A $0$-form $\vph$ on $\man$ is a real valued differentiable function.
One may interpret the form $\vph$ as a field describing a certain
intensive property in $\man$ such as the temperature field, a certain
potential field, etc. A $0$-form does not induce nontrivial hyperplanes
and so no real material structure is represented by $\gf$. In addition,
the condition $\dee\vph=0$ is not really a condition of integrability
as $\vph$ cannot be the exterior derivative of a form. However, the
nonuniformity of $\vph$, implied by $\dee\vph\ne0$ may still be
regarded as a representation of a field of defects. This is manifested
clearer in the case where we consider currents. The currents under
consideration will be of order $n$. 

Consider for example an $n$-dimensional submanifold with boundary
$\body\subset\man$. Let $\curr_{\body}$ be the $n$-current in $\man$
given by
\begin{equation}
\curr_{\body}(\form)=\int_{\body}\form.
\end{equation}
It follows from (\ref{eq:BdryOfCurrIndcdBySubman}) that$\bdry\curr_{\body}(\ga)=\curr_{\bdry\body}(\ga)$.
This identity suggests that the boundaries of bodies be interpreted
as defects. The condition $\bdry^{2}\curr_{\body}=0$ simply imply
in this case that the boundary of $\bdry\body$ vanishes.

\subsection{Schwartz Distributions: The Case of $0$-Currents\label{sec:zero-currents}}

Differential forms of degree $n$ may be integrated over bounded subsets
of the material manifold $\man$. As such, from the physical point
of view, they represent densities of extensive properties such as
the mass density or electric charge density. Such forms may be paired
with smooth $0$-forms of compact supports, that is, with test functions
over $\man$. Thus, if $\rho$ is an $n$-form, one may consider the
$0$-current $\curr_{\rho}$ given by, 
\begin{equation}
\curr_{\rho}(\gf)=\int_{\man}\rho\gf\label{eq:ZeroCurrInducedByN-Forms}
\end{equation}
for every test function $\gf$. Evidently, the product with the test
function $\gf$ cuts off the integrand so that if $\rho$ is measurable,
the integral is well defined. The test function $\gf$ may be interpreted
as a potential so that $\gf\rho$ may be interpreted as the corresponding
energy density. 

An $(n-1)$-form $\st$ induces another construction of a $0$-current
$\bdry\curr_{\st}$ by 
\begin{equation}
\bdry\curr_{\st}(\gf)=\curr_{\st}(\dee\gf)=\int_{\man}\st\wedge\dee\gf.
\end{equation}
It is noted that $\curr_{\st}$ is a $1$-current induced by the form
$\st$. If we interpret the test form $\gf$ as a potential, $\dee\gf$
may be interpreted as (minus) the corresponding force field and $\st$
may be interpreted as the flux field for some extensive property under
consideration, so that $\st\wedge\dee\gf$ is the density of power.

For any $n$-form $\rho$, one has $\dee\rho=0$. In analogy, $0$-currents
have no boundary. Hence, no defects may be associated with such densities.
Nevertheless, we may interpret the $n$-form $\rho$ as the void fraction
or density of vacancies in the body.

Singular $0$-currents are singular distributions defined on the manifold
$\man$. Thus, in addition to currents induced by $n$-forms as in
 (\ref{eq:ZeroCurrInducedByN-Forms}), one may consider distributions
such as the Dirac measure $\gd_{x}$ at a point $x\in\man$ , \ie, the
current defined by
\begin{equation}
\gd_{x}(\gf)=\gf(x)
\end{equation}
for any test function $\gf$. In addition, for a $0$-current $\curr$
and a vector field $\vf$, one may consider the $0$-current $\bdry(\vf\wedge\curr)$
which acts on test functions by 
\begin{equation}
\bdry(\vf\wedge\curr)(\gf)=(\vf\wedge\curr)(\dee\gf)=\curr(\dee\gf(\vf)).
\end{equation}
For example, 
\begin{equation}
\bdry(\vf\wedge\gd_{x})(\gf)=\gd_{x}(\dee\gf(\vf))=(\dee\gf(x))(\vf(x))
\end{equation}
which is the directional derivative of $\gf$ at $x$ in the direction
of $\vf(x)$.

Singular $0$-currents may be interpreted as concentrated vacancies
or inclusions. For example, $(\dee\gf(x))(\vf(x))$ may be interpreted
as the power expended by the force $\dee\gf(x)$ for the velocity
$\vf(x)$ of the concentrated inclusion.

\section{Dislocations\label{sec:Dislocations}}

The description of smooth distributions of dislocations in terms of
differential forms on general manifolds and the generalization to
singular dislocations using de Rham currents are discussed in our
previous paper \cite{ESdislocations2012}. Here, following the general
introduction below and the reviewing the example of an edge dislocation,
we will consider the example of a screw dislocation which we did not
consider in \cite{ESdislocations2012}. Finally, we will demonstrate
how the Frank rules follow from the condition $\bdry^{2}\curr=0$.

\subsection{The Geometry of Dislocations}

Continuous distributions of dislocations in the body $\man$ are
associated with the integrability issue of a $1$-form $\gf$. Each
$1$-form is trivially decomposable and as such, it induces at each
$x\in\man$ a hyperplane $D_{x}$ which we interpret as the Bravais
hyperplane at that point. Multiplying $\gf$ be a positive function
$a$ will have the effect of changing the density of the Bravais hyperplanes.
In fact, the covector $\gf(x)$ is intimately related to the Miller
indices for the Bravais planes at $x$. It is natural therefore to
refer to the $1$-form $\gf$ as the layering form. 

Rather than considering the Burgers vector obtained by tracing a loop
in the non dislocated body and evaluating the vector needed to close
the loop in the dislocated state, we envisage an integration over
a closed loop of the form $\gf$ which is interpreted as the total
amount of hyperplanes that penetrate the loop in one particular orientation.
It is noted that being a $1$-form, the distribution induced by $\gf$
is not necessarily involutive. For a smooth layering $1$-form $\gf$,
the distribution of dislocations is modeled by $\dee\gf$. In case
$\dee\gf=0$, locally, by the Poincare lemma there is a function $u$
such that $\vf=\dee u$. We view $u$ as a labeling function for the
Bravais hypersurfaces. For additional examples to those given below,
see \cite{ESdislocations2012}.

For the singular case, the layering is modeled by an $(n-1)$-structure
current $\curr$ and the dislocations are described by its boundary
$\bdry\curr$.

\subsection{Edge Dislocations}

Assume that $\man$ is an $n$-dimensional manifold without boundary
and let $\hyper$ be an $(n-1)$-submanifold with boundary of $\man$.
We consider the $(n-1)$-structure current $\curr_{\hyper}$ given
by ($\ref{eq:CurrBySubmanifold}$). Then, as shown in (\ref{eq:BdryOfCurrIndcdBySubman}),
the dislocation $(n-2)$-current is given by $\curr_{\bdry\hyper}$.

As a concrete example in $\rthree$, consider the case where $\man=(-1,1)^{3}$
is an open cube in $\rthree$. Let 
\begin{equation}
\hyper=\{(0,x^{2},x^{3})\in\man\mid\, x^{2}\les0\}
\end{equation}
equipped with the orientation induced by the form $\dee x^{2}\wedge\dee x^{3}$.
The current $\curr_{\hyper}$ represents an added half plane. Then,
$\bdry\curr_{\hyper}=\curr_{\bdry\hyper}$, where $\bdry\hyper=\{(0,0,x^{3})\in\man\}$,
oriented naturally by the form $\dee x^{3}$, is the singular dislocation
line as expected.

\subsection{Screw Dislocations\label{sub:Screw-dislocations}}

We present here an additional example, that of a screw dislocation.

Let $L\subset\reals^{3}$ be given by $L=\{(0,0\}\times\reals\}=\{(0,0,z)\mid z\in\reals\}$
and let $D\subset\rthree$ be given by $D=\rthree\setminus L=\{(x,y,z)\in\rthree\mid(x,y)\ne(0,0)\}$.
It is noted that on $D$ we may use a cylindrical coordinate system
$(r,\theta,z)$, where we take the domain $[0,2\pi)$ for $\theta$
without using a proper atlas on the unit circle.

Consider the layering $1$-form $\vph$ on $D$ defined by 
\begin{equation}
\vph=-\frac{b}{2\pi}\dee\theta+\dee z.\label{eq:ScrewForm}
\end{equation}
Evidently, as its components are constants, $\vph$ is a closed form.
It thus follows from Poincare's lemma that locally $\vph$ is exact.
Since $D$ is not contractible to a point, $\vph$ is not exact globally.
In fact, in the open set $D\setminus\{(r,\theta,z)\mid\theta=0\}$,
$\vph=\dee F$ for the real valued 
\begin{equation}
F(r,\theta,z)=-\frac{b\theta}{2\pi}+z
\end{equation}
whose level sets 
\begin{equation}
z=\frac{b\theta}{2\pi}+C,\qquad C\in\reals
\end{equation}
describe spiraling screw threads of pitch $b$.

For any $r>0$, let $S_{r,l}=\{(x,y,z)\in\rthree\mid x^{2}+y^{2}=r^{2},\, z=l\}$
be the circle of radius $r$ situated at $z=l$ and let $\incl:S_{r,l}\to D$
be the inclusion. Then, for example, 
\begin{equation}
\begin{split}\int_{S_{r,l}}\vph & =\int_{S_{r,l}}\incl^{*}(\vph),\\
 & =\int_{S_{r,l}}-\frac{b}{2\pi}\dee\theta,\\
 & =-b.
\end{split}
\end{equation}
(It is observed that $\incl^{*}(\vph)(\bdry/\bdry\theta)=\vph(\incl_{*}(\bdry/\bdry\theta))=\vph(\bdry/\bdry\theta)=-b/2\pi$.)

We now consider the structure $2$-current $\curr_{\vph}$ in $\reals^{3}$,
given by
\begin{equation}
\curr_{\vph}(\form)=\int_{D}\vph\wedge\form\label{eq:FormForScrew}
\end{equation}
for any $2$-form $\form$ on $\rthree$ with compact support. In
order to determine the associated geometry of the dislocation, we
examine the defect current, the boundary $\bdry\curr_{\vph}$. For
any $1$-form $\ga$, we have,
\begin{equation}
\begin{split}\bdry\curr_{\vph}(\ga) & =\curr_{\vph}(\dee\ga),\\
 & =\int_{D}\vph\wedge\dee\ga,\\
 & =-\int_{D}\dee(\vph\wedge\ga)+\int_{D}\dee\vph\wedge\ga.
\end{split}
\end{equation}
Since $\dee\vph=0$ in $D$, we conclude that 
\begin{equation}
\bdry\curr_{\vph}(\ga)=-\int_{D}\dee(\vph\wedge\ga).
\end{equation}

Let $C_{\eps}=\{(x,y,z,)\in\rthree\mid x^{2}+y^{2}<\eps^{2}\}$ and
let $D_{\eps}=\rthree\setminus C_{\eps}$. We may write
\begin{equation}
\bdry\curr_{\vph}(\ga)=-\int_{D}\dee(\vph\wedge\ga)=-\lim_{\eps\to0}\int_{D_{\eps}}\dee(\vph\wedge\ga).\label{eq:screw}
\end{equation}
Now it is noted that $D_{\eps}$ is a manifold with a boundary. In
fact, setting $S_{\eps}=\{(x,y)\in\reals^{2}\mid x^{2}+y^{2}=\eps^{2}\}$,
$\bdry D_{\eps}=S_{\eps}\times\reals$. We may therefore use Stokes's
theorem in  (\ref{eq:screw}) and obtain
\begin{equation}
\bdry\curr_{\vph}(\ga)=-\lim_{\eps\to0}\int_{\bdry D_{\eps}}\incl^{*}(\vph\wedge\ga),\label{eq:screw1}
\end{equation}
where $\incl^{*}(\vph\wedge\ga)$ is the pullback under the inclusion
$\incl:\bdry D_{\eps}\to D_{\eps}$ which is simply the restriction
of $\vph\wedge\ga$ to vectors tangent to $\bdry D_{\eps}$.

A $1$-form $\ga$ is represented by $\ga=\ga_{x}\dee x+\ga_{y}\dee y+\ga_{z}\dee z$
for the smooth functions $\ga_{x}$, $\ga_{y}$, and $\ga_{z}$ defined
on $\rthree$. In $D$, the form $\ga$ may also be represented using
cylindrical coordinates as $\ga=\ga_{r}\dee r+\ga_{\theta}\dee\theta+\ga_{z}\dee z$.
Since $\ga_{x}\dee x+\ga_{y}\dee y=\ga_{r}\dee r+\ga_{\theta}\dee\theta$,
using $x=r\cos\theta$, $y=r\sin\theta$ and 
\begin{equation}
\dee x=\parby xr\dee r+\parby x{\theta}\dee\theta,\quad\dee y=\parby yr\dee r+\parby y{\theta}\dee\theta,
\end{equation}
one has
\begin{equation}
\ga_{\theta}=r(-\ga_{x}\sin\theta+\ga_{y}\cos\theta).
\end{equation}
The restriction to $\bdry D_{\eps}$ satisfies
\begin{equation}
\incl^{*}(\vph\wedge\ga)=(\vph_{\theta}\ga_{z}-\vph_{z}\ga_{\theta})\dee\theta\wedge\dee z=\left(-\frac{b}{2\pi}\ga_{z}-\ga_{\theta}\right)\dee\theta\wedge\dee z,
\end{equation}
and it follows that
\begin{equation}
\begin{split}\bdry\curr_{\vph}(\ga) & =\lim_{\eps\to0}\int_{-\infty}^{\infty}\dee z\left[\int_{S_{\eps}}\left(\frac{b}{2\pi}\ga_{z}+\ga_{\theta}\right)\dee\theta\right],\\
 & =\int_{-\infty}^{\infty}\dee z\left\{ \lim_{\eps\to0}\left[\int_{S_{\eps}}\left(\frac{b}{2\pi}\ga_{z}-+\ga_{\theta}\right)\dee\theta\right]\right\} .
\end{split}
\label{eq:screw2}
\end{equation}
Examining the limit in the second line of  (\ref{eq:screw2}), we
first note that
\begin{equation}
\begin{split}\lim_{\eps\to0}\int_{S_{\eps}}\ga_{\theta}\dee\theta & =\lim_{\eps\to0}\int_{S_{\eps}}\eps(-\ga_{x}\sin\theta+\ga_{y}\cos\theta)\dee\theta,\\
 & =0,
\end{split}
\end{equation}
since $\ga_{x}\to\ga_{x}(x=0,y=0,z)$, $\ga_{y}\to\ga_{y}(x=0,y=0,z)$,
as $\eps\to0$ (and thus are independent of $\theta$), and since
the integrals of the trigonometric functions of over the circle vanish.
In addition,
\begin{equation}
\lim_{\eps\to0}\int_{S_{\eps}}\frac{b}{2\pi}\ga_{z}\dee\theta=b\ga_{z}(0,0,z),
\end{equation}
and one concludes that
\begin{equation}
\bdry\curr_{\vph}(\ga)=b\int_{-\infty}^{\infty}\ga_{z}(0,0,z)\dee z.
\end{equation}
If we assign the natural orientation to $L=\{(0,0)\}\times\reals\subset\rthree$,
we may use $\curr_{L}$ to denote the $1$-current given by 
\begin{equation}
T_{L}(\ga)=\int_{L}\incl_{L}^{*}(\ga).
\end{equation}
Here $\incl_{L}^{*}:L\to\rthree$ is the natural inclusion so that
for any $1$-form $\ga=\ga_{x}\dee x+\ga_{y}\dee y+\ga_{z}\dee z$,
$\incl_{L}^{*}(\ga)=\ga_{z}\dee z$. Thus, we may write the current
as
\begin{equation}
\bdry\curr_{\vph}=bT_{L}.
\end{equation}

\begin{rem}
Using the same notation as above, consider the case where instead
of $\vph$ given in (\ref{eq:ScrewForm}) one has the $1$-form $\vph'$
given by 
\begin{equation}
\end{equation}
\begin{equation}
\vph'=-\frac{b}{2\pi}\dee\theta.
\end{equation}
Since $\vph'$ is annihilated by the vector space spanned by the base
vectors $\bdry/\bdry r$ and $\bdry/\bdry z$, the layers induced
by $\vph'$ look like the pages of a book spread evenly in all directions.
If we follow the same steps as above we obtain 
\begin{equation}
\vph'\wedge\ga=-\frac{b}{2\pi}(\ga_{r}\dee\theta\wedge\dee r+\ga_{z}\dee\theta\wedge\dee z),
\end{equation}
so that
\begin{equation}
\incl^{*}(\vph'\wedge\ga)=-\frac{b}{2\pi}\ga_{z}\dee\theta\wedge\dee z.
\end{equation}
If follows that $\bdry\curr_{\vph'}=\bdry\curr_{\vph}$. This observation
may be viewed as follows. Let $\curr_{\dee z}$ be the current induced
by the form $\dee z$. Then, since $\dee^{2}z=0$, $\bdry\curr_{\dee z}=0$.
Since $\curr_{\vph}=\curr_{\vph'}-(b/2\pi)\curr_{\dee z}$, it follows
that $\bdry\curr_{\vph}=\bdry\curr_{\vph'}$. Alternatively, one may
envisage a smooth twist of $\rthree$ about the $z$-axis under which
the book is deformed into the book. Since our objects are invariant
under diffeomorphisms, both layering structures have the same dislocations.
Thus for example, a similar observation will hold if the pages of
the book are not plane but are bent perpendicularly to the $z$-axis
forming the shape of a whirlpool.
\end{rem}

\subsection{The Frank Rules for Dislocations}

If the $2$-form $\psi$ describes the continuous distribution of
dislocations, the Frank rules are induced by the compatibility condition
$\dee\psi=0$. For the singular case, if an $(n-2)$-current $R$
represents the geometry of the dislocations, the compatibility condition
that induces Frank's rules is $\bdry R=0$. 

For example, let $\hyper$ be an $(n-1)$-dimensional submanifold
with boundary of $\man$ and consider the the $(n-2)$-current $R$
so that
\begin{equation}
R(\psi)=\int_{\bdry\hyper}u\psi
\end{equation}
for some given differentiable function $u$ defined on $\bdry\hyper$.
then, the boundary $\bdry R$ is given by 
\begin{equation}
\begin{split}\bdry R(\ga) & =\int_{\bdry\hyper}u\dee\ga,\\
 & =\int_{\bdry\hyper}\dee(u\ga)-\int_{\bdry\hyper}\dee u\wedge\ga,\\
 & =\int_{\bdry^{2}\hyper}u\ga-\int_{\bdry\hyper}\dee u\wedge\ga,\\
 & =-\int_{\bdry\hyper}\dee u\wedge\ga.
\end{split}
\end{equation}
Here, $\ga$ is any $(n-3)$-form so for the three dimensional case
$\ga$ is any smooth function of compact support. Since compatibility
imposes the condition $\bdry R=0$, it follows that for $R$ to be
a dislocation current, the function $u$ must be constant on $\bdry S$
which is Frank's first rule.

\section{Inclinations and Disclinations\label{sec:Disclinations}}

Disclinations are viewed here as defects in the arrangements of $1$-dimensional
subspaces, or directors. As in \cite{FrankLiquidCrystals} and \cite{Chandrasekhar},
 this field may indicate the inclinations of the optical axes of
liquid crystals. The interpretation of disclinations as defects in
the orientations of the Bravais planes (e.g., \cite{kronerAnthony75})
may be viewed in some cases as defects in the arrangements of the
normal vectors to the respective Bravais planes. Such cases can be
described using the framework outlined below.

Thus, disclinations are represented as boundaries of currents of order
$1$. In the smooth case such a current is represented by an $(n-1)$-form
$\vph$, the inclination form, and the structure of the disclinations
is given by the $n$-form $\dee\vph$. It is noted that any $(n-1)$-form
is decomposable. (See \cite[Section 1.V]{Sternberg}, and \cite{Worldlines}
for a continuum mechanical application.) The induced distribution
is necessarily involutive and the $1$-dimensional integral submanifolds
to which the directors are tangent may be easily constructed as follows.

At each point $x\in\man$ where $\vph(x)\ne0$, $\vph(x)$ determines
a unique 1-dimensional subspace $\vs_{x}$ of the tangent space $T_{x}\man$
by $v\contr\vph(x)=0$ for each $v\in\vs_{x}$. The collection of
subspaces $\vs_{x}$ forms a $1$-dimensional distribution. The $1$-dimensional
subspace $\vs_{x}$ may be determined as follows. Let $\vel$ be a
volume element on $\man$. Locally, $\vel$ may be represented in
the form 
\begin{equation}
\vel=\vel_{0}\lisupp{\dee x}1{\wedge}n
\end{equation}
for a positive real valued function $\vel_{0}$ and $\vph$ may be
represented locally in the form
\[
\vph=\sum_{i=1}^{n}\vph_{1\dots\hat{\imath}\dots n}\lisupwout{\dee x}{\wedge}ni
\]
where a ``hat'' indicates the omission of an element. Then, there
is a unique tangent vector $u$ such that $u\contr\vel=\vph$. If
a vector $u$ is represented by $u=\sum_{i}u^{i}\bdry/\bdry x^{i}$,
then, $u\contr\vel$ is represented by 
\begin{equation}
u\contr\vel=\sum_{i=1}^{n}\sgn{i-1}\vel_{0}u^{i}\lisupwout{\dee x}{\wedge}ni.
\end{equation}
Thus, as $\vel_{0}\ne0$, there is always a vector field $u$ satisfying
$u\contr\vel=\vph$ and its components are given locally by 
\begin{equation}
u^{i}=\sgn{i-1}\frac{\vph_{1\dots\hat{\imath}\dots n}}{\vel_{0}}.\label{eq:vectorsFieldsAndVolumeElements}
\end{equation}
If we select a different volume element, the only parameter that will
change in the equation above will be the positive number $\vel_{0}$
and so the resulting vector will be in the same one dimensional subspace.
Thus, the form $\vph$ determines a unique oriented 1-dimensional
subspace $\mathbf{U}_{x}$ at each $x$ such that $\vph(x)\ne0$.
If no particular orientation is chosen on $\man$ no orientation will
be induced on $\mathbf{U}_{x}$. The space $\vs_{x}$ and $\mathbf{U}_{x}$
are isomorphic. Let $\vel$ be a volume elements and $u$ the vector
such that $\vph=u\contr\vel$. Then, any nonzero $v\in\mathbf{U}_{x}$
is of the form $v=au$, $a\ne0$. Thus, $v\contr(u\contr\vel)=au\contr(u\contr\vel)=0$,
because $\vel(u,u,v_{3},\dots,v_{n})=0$ for any collection of vectors
$v_{3},\dots,v_{n}$.  

For an $(n-1)$-form $\vph$ we interpret the distribution $\vs$
of 1-dimensional subspaces of the tangent space as indicating the
inclinations of the directors in the body. Multiplying the form $\vph$
by a positive number will affect the ``density'' of the directors.

Unlike the case of Bravais hyperplanes, inclination fields are always
involutive, \ie, at each point $x\in\man$ there is a curve $c_{x}:(-\eps,\eps)\to\man$,
$\eps>0$, such that $c_{x}(0)=x$ and the tangent vector to the curve
satisfies 
\begin{equation}
\left.\ddt{c_{x}}\right|_{t=0}\in\vs_{x}.
\end{equation}

Since we have assumed that the form $\vph$ is differentiable, it
follows that for a choice of a smooth volume element $\vel$, the
representing vector field $u$ is differentiable. Hence, the theorems
on the existence and uniqueness of the solutions of ordinary differential
equations imply the existence of the integral lines to the vector
field $u$, \ie, at each point $x\in\man$ there is a curve $c_{x}:(-\eps,\eps)\to\man$,
$\eps>0$, such that $c_{x}(0)=x$ and the tangent vector to the curve
\begin{equation}
\left.\ddt{c_{x}}\right|_{t=0}=u(x).
\end{equation}

An inclination form may be integrated over $(n-1)$-dimensional submanifolds
of $\man$. Let $\srfc$ be an oriented $(n-1)$-dimensional submanifold
of $\man$. Then, 
\begin{equation}
\Phi_{\srfc}=\int_{\srfc}\vph\label{eq:BdryIntegralDirectors}
\end{equation}
is interpreted as the total amount of directors penetrating the surface
$\srfc$. It should be noted that $\Phi_{\srfc}$ depends on the orientation
of $\srfc$ and that the restriction of $\vph$ to a point in $\srfc$
may be of the same orientation as $\srfc$ or the inverse orientation.
Thus, for a nonvanishing inclination form, the total $\Phi_{\srfc}$
may vanish which implies that each of the integral lines penetrates
$\srfc$ in one orientation the same number of times that it penetrates
$\srfc$ in the opposite orientation.

For the inclination $(n-1)$-form $\vph$, the distribution of smooth
disclinations induced is the exterior derivative, the $n$-form $\dee\vph$.
Thus, for a $n$-dimensional submanifold with boundary $\body\subset\man$,
letting $\hyper=\body$ in (\ref{eq:BdryIntegralDirectors}), $\Phi_{\bdry\body}$
is interpreted as the total amount of directors that penetrate $\bdry\body$.

Stokes's theorem implies immediately that
\begin{equation}
\Phi_{\bdry\body}=\int_{\body}\dee\vph,
\end{equation}
so that $\Phi_{\bdry\body}$ is the integral of the disclination field
over $\body$. Figuratively speaking, the disclination field represents
the source term for the directors.

 It is observed that for any given vector field one can label the
integral lines by a submanifold of dimension $n-1$ of initial conditions
(see \cite[pp.~246--247]{AbeMarsdenManifolds}). However, the vector
fields induced by $\gf$ depend on the choice of volume element $\theta$.
Thus, such labeling is not unique and the presence of disclinations
will reflected by $\dee\gf$.

An inclination $(n-1)$-form $\vph$ induces a de Rham $1$-current
$\curr_{\vph}$ as in (\ref{eq:SmoothCurr}).  In the non-smooth
case, we replace the inclination $1$-form $\vph$ and the current
it induces by a general \emph{inclination $1$-current} $T$. Inclination
currents that are not given in terms smooth $(n-1)$-forms represent
singular, or concentrated, director fields as the examples below illustrate.

\begin{example}
\textbf{A non-coherent interface 1.} Consider the locally integrable
$(n-1)$-form $\vph$ in $\rn$ given by 
\begin{equation}
\vph(x)=\begin{cases}
\lisupc{\dee x}{\wedge}{n-1}, & \text{for }x\in\cprn,\\
a\lisupc{\dee x}{\wedge}{n-1}, & \text{for }x\in\nrn,
\end{cases}
\end{equation}
where $a\in\reals$ is a constant, $\nrn=\{x\in\rn\mid x^{n}<0\}$,
and $\cprn=\{x\in\rn\mid x^{n}\ges0\}$. The inclination form $\vph$
induces a $1$-current $\curr_{\vph}$ by
\begin{equation}
\curr_{\vph}(\go)=\int_{\rn}\vph\wedge\go.
\end{equation}
Clearly, the $1$-dimensional subspace spanned by $\bdry/\bdry x^{n}$
annihilates $\vph(x)$ for all $x$ for which $x^{n}\ne0$. Thus,
the directors are aligned in the $x^{n}$ direction.

For any smooth compactly supported $0$-form $\ga$ in $\rn$,
\begin{equation}
\begin{split}\bdry\curr_{\vph}(\ga) & =\int_{\rn}\vph\wedge\dee\ga,\\
 & =\int_{\nrn}\vph\wedge\dee\ga+\int_{\cprn}\vph\wedge\dee\ga,\\
 & =(-1)^{n-1}\left[\int_{\nrn}\dee(\ga\vph)-\int_{\nrn}\ga\dee\vph+\int_{\prn}\dee(\ga\vph)-\int_{\prn}\ga\dee\vph\right],\\
 & =(-1)^{n-1}\left[\int_{\bdry\nrn}\ga\vph+\int_{\bdry\prn}\ga\vph\right]
\end{split}
\label{eq:IncoherenceComputations}
\end{equation}
where in the third line we used  (\ref{eq:ExterDerOfExterProd}).
Let $P$ be the hyperplane in $\rn$ defined by $x^{n}=0$ oriented
such that $P=\bdry\nrn=-\bdry\prn$ so that $\vel_{P}=\lisupc{\dee x}{\wedge}{n-1}$
is the natural volume element on $P$. Let $\curr_{P}$ be the $0$-current
given by
\begin{equation}
\curr_{P}(\ga)=\int_{P}\ga\vel_{P}.
\end{equation}
We conclude that 
\begin{equation}
\bdry\curr_{\vph}=(-1)^{n-1}(a-1)\curr_{P},
\end{equation}
which is interpreted as a concentrated source of directors of magnitude
$a-1$ which is distributed over the $\lisup x,{n-1}$ hyperplane.
\end{example}

\begin{example}
\textbf{A non-coherent interface 2.} Consider the locally integrable
1-form $\vph$ in $\rn$ given by 
\begin{equation}
\vph(x)=\begin{cases}
\vph_{2}, & \text{for }x\in\cprn,\\
\vph_{1}, & \text{for }x\in\nrn,
\end{cases}
\end{equation}
where $\vph_{1}$ and $\vph_{2}$ are uniform $(n-1)$-forms in $\nrn$
and $\cprn$, respectively. Letting $\curr_{\vph}$ be the $1$-current
defined by
\begin{equation}
\curr_{\vph}(\go)=\int_{\rn}\vph\wedge\go,
\end{equation}
it follows from  (\ref{eq:IncoherenceComputations}) that
\begin{equation}
\bdry\curr_{\vph}(\ga)=\int_{P}\ga(\vph_{1}-\vph_{2}).
\end{equation}
We conclude that the disclination current vanishes if $\vph_{1}$
and $\vph_{2}$ have the same restriction to $P$, \ie, both forms
have the same component relative to $\lisupc{\dee x}{\wedge}{n-1}$.
In particular, let $v$ be a vector parallel to the $\lisup x,{n-1}$-plane.
Then, $\vph_{2}=v\contr\lisupc{\dee x}{\wedge}n$ is annihilated by
the $1$-dimensional space spanned by $v$ and all the components
of $\vph_{2}$ that do not vanish correspond to basis elements of
the form 
\[
\lisuppwout{\dee x}1{\wedge}{\cdots}nk,\quad k=1,\dots,n-1.
\]
In this case, the directors corresponding the $\vph_{2}$ do not intersect
the $\lisup x,{n-1}$-plane, the component of $\vph_{2}$ relative
to $\lisupc{\dee x}{\wedge}{n-1}$ vanishes, and 
\begin{equation}
\bdry\curr_{\vph}(\ga)=\int_{P}\ga\vph_{1}.
\end{equation}

\end{example}

\begin{example}
\textbf{An edge disclination.} Let $L$ be a connected and oriented
$1$-dimensional submanifold with a boundary of $\man$. Then, $L$
induces a $1$-current $\curr_{L}$ by
\begin{equation}
\curr_{L}(\form)=\int_{L}\form,
\end{equation}
for all compactly supported smooth $1$-forms $\form$ in $\man$.
Using Stokes's theorem, one has
\begin{equation}
\bdry\curr_{L}(\ga)=\int_{L}\dee\ga=\int_{\bdry L}\ga.
\end{equation}
Evidently, as $\bdry L$ is a $0$-dimensional submanifold, and assuming
it is not empty, it may contain one or two points, each having either
a positive or a negative orientation while the other point, if exists,
has the opposite orientation.

In the case where $\bdry L$ contains one point $x_{1}$ and assuming
its orientation is positive, one has $\bdry L(\ga)=\ga(x_{0})$, representing
an edge disclination originating at $x_{0}$. This will be the situation
if $\man=(-1,1)^{3}\subset\rthree$ and $L=\{(0,0,z)\mid-1<z\les0\}$
so that $x_{1}=(0,0,0)$. In this case the disclination does not terminate
inside the body. In the case where $\bdry L$ contains also the additional
point $x_{2}$ having a negative orientation, $\bdry L(\ga)=\ga(x_{1})-\ga(x_{2})$
and the disclination terminates at $x_{2}$.
\end{example}

\begin{example}
\textbf{Directors emanating from a singular line.} Using the notation
introduced in Section \ref{sub:Screw-dislocations} on screw dislocations,
consider the inclination $n-1=2$-form $\vph$ defined on $D\subset\rthree$
by 
\begin{equation}
\vph=\dee\theta\wedge\dee z.
\end{equation}
The inclination form induces an inclination $1$-current $\curr$
on $\rthree$ by the right had side of (\ref{eq:FormForScrew}). It
is noted that in its domain of definition, $\dee\vph=0$.

To compute the disclination $0$-current $\bdry\curr$, one observes
that for any smooth function $\ga$, compactly supported in $\rthree$,
 (\ref{eq:ExterDerOfExterProd}) implies that 
\begin{equation}
\begin{split}\bdry\curr(\ga) & =\int_{D}\dee(\ga\vph)-\int_{D}\ga\dee\vph,\\
 & =\lim_{\eps\to0}\int_{D_{\eps}}\dee(\ga\vph),\\
 & =\lim_{\eps\to0}\int_{\bdry D_{\eps}}\incl^{*}(\ga\vph),\\
 & =\lim_{\eps\to0}\int_{\bdry D_{\eps}}\ga\dee\theta\wedge\dee z.
\end{split}
\end{equation}
In analogy with the computations of Section \ref{sub:Screw-dislocations},
one obtains
\begin{equation}
\bdry\curr(\ga)=2\pi\int_{z=-\infty}^{\infty}\ga(0,0,z)\dee z,
\end{equation}
which we may write as
\begin{equation}
\bdry\curr=2\pi T_{L}\fcontr\dee z.
\end{equation}
Thus, we have a uniform distribution of directors' source along the
$z$-axis.
\end{example}

\section{Kinematics of Defect Distributions\label{sec:Kinematics}}

In this section we consider the kinematics of the material structure
and the distribution of defects. Noting that material structure and
the associated defects are viewed here as intrinsic to a body and
unrelated to the kinematics of the body in space, in the following
two subsections we consider the motion of material structure and defects
resulting from a family of diffeomorphisms of the body. (See \cite{FSreynoldsSmoothDef2013}
for another application of the same mathematical notions.) In other
words, the material structure, as represented by a smooth form and
its exterior derivative or a de Rham current and its boundary, are
carried with material diffeomorphisms. In contrast, the last subsection
proposes an example for an evolution of a continuously distributed
material structure to a singular one using a process which is the
opposite of smoothing.

\subsection{Smooth Evolutions of Structure Forms and Continuously Distributed
Defects}

In order to study the deformation of structure forms and currents,
we consider the following setting. It is assumed that we are given
a time dependent flow, or a smooth evolution operator, $\flow:\inter^{2}\times\man\to\man$
in the interval $\inter=[a,b]\subset\reals$. That is, for each time
instances $t$ and $\tau$, with $t\les\tau$, $t,\tau\in\inter$,
$\flow_{\tau,t}:\man\to\man$ is a diffeomorphism and $\flow_{\tau,t}\comp\flow_{t,s}=\flow_{\tau,s}$.
For $t\ges\tau$, $\Phi_{\tau,t}=\Phi_{t,\tau}^{-1}$, which implies
that $\Phi_{t,t}=I_{\man}$, the identity diffeomorphism. Evidently,
the flow induces a smooth homotopy $h:[a,b]\times\man\to\man$ by
$h(t,x)=h_{t}(x)=\flow_{t,a}(x)$ so that $\flow_{\tau,t}=h_{\tau}\comp h_{t}^{-1}$.
The time dependent flow induces a time dependent vector field $\vf:\inter\times\man\to T\man$
by setting
\begin{equation}
\vf(t,x)=\resto{\frac{\partial}{\partial\tau}}{\tau=t}\flow_{\tau,t}(x)=\resto{\frac{\partial}{\partial\tau}}{\tau=t}h_{\tau}(h_{t}^{-1}(x)),
\end{equation}
that is, $\vf(t,x)$ is the tangent at the time $\tau=t$ to the curve
\begin{equation}
c_{t,x}(\tau)=\flow_{\tau,t}(x)=h_{\tau}(h_{t}^{-1}(x)),
\end{equation}
starting at $x$ at time $t$ (see for example \cite[p.~283]{AbeMarsdenManifolds}).
Conversely, the flow is the solution of the differential equation
\begin{equation}
\vf(s,\flow_{s,t}(x))=\resto{\frac{\partial}{\partial\tau}}{\tau=s}c_{t,x}(\tau)=\resto{\frac{\partial}{\partial\tau}}{\tau=s}\flow_{\tau,t}(x)=\resto{\frac{\partial}{\partial\tau}}{\tau=s}h_{\tau}(h_{t}^{-1}(x)).\label{eq:VectorFieldAssocWithAFlow}
\end{equation}
Alternatively, setting $t=a$,  the differential equation may be
expressed as 
\begin{equation}
\vf(s,h_{s}(x))=\resto{\frac{\partial}{\partial\tau}}{\tau=s}h_{\tau}(x).
\end{equation}

Each diffeomorphism $\flow_{\tau,t}$ induces the pullback of forms
$\flow_{\tau,t}^{*}$ from $\image h_{\tau}$ onto $\image h_{t}$.
In particular, if $\form$ is an $r$-form with compact support in
$\man$, the same holds for its pullback $\flow_{\tau,t}^{*}\form$. 

Let $\gf$ be a (time independent) differentiable, material structure
$(n-r)$-form. Then, for each $t\in\inter$, the flow induces a time
dependent $(n-r)$-form $\flow_{\tau,t}^{*}\gf$ and in particular
the form $h_{t}^{*}\gf$. It is also recalled that for any $(n-r$)
-dimensional submanifold $\srfc\subset\man$, and a form $\gf$ having
a compact support, one has \cite[p.~466]{AbeMarsdenManifolds}
\begin{equation}
\int_{\srfc}h_{t}^{*}\gf=\int_{h_{t}\{\srfc\}}\gf.
\end{equation}
This will hold in the particular case where the submanifold $\srfc$
is compact so that the restriction of $h_{t}^{*}\gf$ to $\srfc$
has a compact support. It is evident from the above observations that
rather than $h_{t}^{*}\gf$, it is the pushforward $h_{t}^{-*}\gf:=h_{t}^{*-1}\gf=h_{t}^{-1*}\gf$
that represents the evolution of the structure form. Specifically,
replacing $h_{t}$ by $h_{t}^{-1}$ and $\srfc$ by $h_{t}\{\srfc\}$
above, one has
\begin{equation}
\int_{h_{t}\{\srfc\}}h_{t}^{-*}\gf=\int_{\srfc}\gf,
\end{equation}
which indicates how the evolution of the structure form follows the
evolution $h_{t}\{\srfc\}$ of the submanifold. It is emmphasized
that we regard the flow $\flow$ to be associated with the structure
of matter only and has nothing to do with the motion in space of the
material points belonging to the body.

It is recalled that for each differentiable mapping $f:\man\to\mathcal{N}$
between a manifold $\man$ and a manifold $\mathcal{N}$, and a differentiable
form $\gf$, one has $f^{*}(\dee\gf)=\dee(f^{*}\gf)$. Thus, in our
setting, $h_{t}^{-*}(\dee\gf)=\dee(h_{t}^{-*}\gf)$, that is, the
smooth field of defects induced by $h_{t}^{-*}\gf$ is obtained by
the pushforward of the field of defects induced by $\gf$. 

We also note that as $f^{*}(\ga\wedge\gb)=f^{*}\ga\wedge f^{*}\gb$
\cite[p.~420]{AbeMarsdenManifolds}. Thus, if $\curr_{\gf}$ is the
$r$-current on $\man$ induced by $\gf$, then for any $r$-form
$\form$ having a compact support on $\man$,
\begin{equation}
\begin{split}\curr_{\gf}(\form) & =\int_{\man}\gf\wedge\form,\\
 & =\int_{\man}h_{t}^{-*}(\gf\wedge\form),\\
 & =\int_{\man}h_{t}^{-*}\gf\wedge h_{t}^{-*}\form.
\end{split}
\end{equation}
It is concluded therefore that 
\begin{equation}
\curr_{\gf}(h_{t}^{*}\form)=\curr_{h_{t}^{-*}\gf}(\form).\label{eq:PushfwdOfCurrInTermsOfForms}
\end{equation}

Next, we would like to compute the rate at which the structure form
evolves under the flow. In general, using the Lie derivative $\lie_{\vf_{\tau}}\form$
of a form $\form$ relative to the vector field $\vf_{\tau}$ associated
with the flow $\flow_{\tau,t}$, one has \cite[p.~372]{AbeMarsdenManifolds}
\begin{equation}
\resto{\parder[\tau]{}}{\tau=s}\flow_{\tau,t}^{*}\form=\flow_{s,t}^{*}(\lie_{\vf_{s}}\form).\label{eq:flowAndLieDerivative}
\end{equation}
In particular, for $s=t$ and for $t=a$, the relation above specializes
to 
\begin{equation}
\resto{\parder[\tau]{}}{\tau=t}\flow_{\tau,t}^{*}\form=\lie_{\vf_{t}}\form,\qquad\resto{\parder[\tau]{}}{\tau=s}h_{\tau}^{*}\form=h_{s}^{*}(\lie_{\vf_{s}}\form).\label{eq:homotopyAndLieDerivative}
\end{equation}
It is noted that the last two equations hold pointwise. It follows
that for each $x\in\man$, 
\begin{equation}
\flow_{\tau_{2},t}^{*}\form(x)-\flow_{\tau_{1},t}^{*}\form(x)=\int_{\tau_{1}}^{\tau_{2}}\flow_{\tau,t}^{*}(\lie_{\vf_{\tau}}\form)(x)\dee\tau.\label{eq:homotopyOfFormsIntOfLie}
\end{equation}

For the rate of change of the pushforward of the structure form, one
has to use in the equations above the time dependent vector field
$\vf^{-1}$ associated with the flow $\flow_{\tau,t}^{-1}$. According
to (\ref{eq:VectorFieldAssocWithAFlow}), it is given by
\begin{equation}
\vf^{-1}(s,\flow_{s,t}^{-1}(x))=\resto{\frac{\partial}{\partial\tau}}{\tau=s}\flow_{\tau,t}^{-1}(x)=\resto{\frac{\partial}{\partial\tau}}{\tau=s}h_{t}(h_{\tau}^{-1}(x)).
\end{equation}
 Since $\flow_{\tau,t}^{-1}\comp\flow_{\tau,t}$ is the identity on
$\man$, 
\begin{equation}
\begin{split}0 & =\resto{\parder[\tau]{}}{\tau=s}\flow_{\tau,t}^{-1}(\flow_{\tau,t}(x)),\\
 & =\resto{\parder[\tau]{}}{\tau=s}\flow_{\tau,t}^{-1}(\flow_{s,t}(x))+T\flow_{s,t}^{-1}\left(\resto{\parder[\tau]{}}{\tau=s}\flow_{\tau,t}(x)\right),\\
 & =\vf^{-1}(s,\flow_{s,t}^{-1}(\flow_{s,t}(x)))+T\flow_{s,t}^{-1}(\vf(s,\flow_{s,t}(x))).
\end{split}
\end{equation}
It is implied that
\begin{equation}
\vf^{-1}(s,x)=-T\flow_{s,t}^{-1}(\vf(s,\flow_{s,t}(x))),
\end{equation}
or,
\begin{equation}
\vf_{s}^{-1}=-T\flow_{s,t}^{-1}\comp(\vf_{s}\comp\flow_{s,t}),
\end{equation}
and in particular,
\begin{equation}
\vf^{-1}(t,x)=-w(t,x).
\end{equation}

The rate of in which the structure form $\gf$ evolves is therefore
\begin{equation}
\begin{split}\resto{\parder[\tau]{}}{\tau=s}\flow_{\tau,t}^{-*}(\gf) & =\flow_{s,t}^{-*}(\lie_{\vf_{s}^{-1}}\gf),\\
 & =-\flow_{s,t}^{-*}(\lie_{T\flow_{s,t}^{-1}\comp(\vf_{s}\comp\flow_{s,t})}\gf).
\end{split}
\end{equation}
We also recall \cite[p.~361]{AbeMarsdenManifolds} that in general,
for a diffeomorphism $f:\man\to\mathcal{N}$, a vector field $\vf$
and a form $\form$, 
\begin{equation}
f^{*}(\lie_{Tf(\vf)}\form)=\lie_{\vf}f^{*}\form,.
\end{equation}
Substituting $\flow_{\tau,t}^{-1}$ for $f$, one has
\begin{equation}
\resto{\parder[\tau]{}}{\tau=s}\flow_{\tau,t}^{-*}(\gf)=-\lie_{\vf_{s}}(\flow_{s,t}^{-*}\gf).
\end{equation}
In particular,
\begin{equation}
\resto{\parder[\tau]{}}{\tau=t}\flow_{\tau,t}^{-*}(\gf)=-\lie_{\vf_{t}}\gf
\end{equation}
is the rate in which the structure form evolves.

Since the Lie derivative commutes with exterior differentiation \cite[p.~428]{AbeMarsdenManifolds},
the rate of change of the distribution of defects is the exterior
derivative of the rate of change of the structure form, \ie,  
\begin{equation}
\resto{\parder[\tau]{}}{\tau=t}\flow_{\tau,t}^{-*}(\dee\gf)=-\lie_{\vf_{t}}\dee\gf=-\dee\lie_{\vf_{t}}\gf=\dee\left(\resto{\parder[\tau]{}}{\tau=t}\flow_{\tau,t}^{-*}(\gf)\right).
\end{equation}

\subsection{Evolutions of General Structure Currents and Defects}

We wish to extend the kinematic analysis for smooth deformations of
structure forms and continuously distributed defects to general, possibly
singular currents. The way this is done is suggested by  (\ref{eq:PushfwdOfCurrInTermsOfForms})
where we observed that for a current $\curr_{\gf}$ induced by a smooth
form $\gf$, the current induced by the evolving form $h_{t}^{-*}\gf$
satisfies $\curr_{h_{t}^{-*}\gf}(\form)=\curr_{\gf}(h_{t}^{*}\form)$. 

Since for each time $t$, $h_{t}$ is a diffeomorphism of $\man$,
given any smooth form $\form$ having a compact support in $\man$,
the pullback $h_{t}^{*}\form$ has a compact support in $\man$, also.
In fact, $h_{t}^{*}$ is a continuous, linear operator on the space
of smooth forms with compact supports in $\man$. Thus, for a diffeomorphism
$f$, the dual operator, the pushforward of currents (or images of
currents \cite[p.~47]{derham}), $f_{*}$, is defined by
\begin{equation}
(f_{*}\curr)(\form)=\curr(f^{*}\form).
\end{equation}
Thus, the evolution of a structure current $\curr$ under the flow
is described by the evolution $h_{t*}\curr$ for which the analysis
above is a special case.

It is observed that
\begin{equation}
\begin{split}\bdry(h_{t*}\curr)(\psi) & =h_{t*}\curr(\dee\psi),\\
 & =\curr(h_{t}^{*}(\dee\psi)),\\
 & =\curr(\dee(h_{t}^{*}\psi)),\\
 & =\bdry\curr(h_{t}^{*}\psi),\\
 & =(h_{t*}(\bdry\curr))(\psi),
\end{split}
\end{equation}
and so,
\begin{equation}
\bdry(h_{t*}\curr)=h_{t*}(\bdry\curr).
\end{equation}
We conclude that the evolution of the defects follows the evolution
of structure current, consistently.

To present a typical example for the evolution of a current which
is not induced by a smooth structure form, consider the $r$-current
$\curr_{\srfc}$ induced by an $r$-dimensional submanifold with boundary
$\srfc$ of $\man$ as in (\ref{eq:CurrBySubmanifold}). Then, 
\begin{equation}
\begin{split}h_{t*}\curr_{\srfc}(\form) & =\curr_{\srfc}(h_{t}^{*}\form),\\
 & =\int_{\srfc}h_{t}^{*}\form,\\
 & =\int_{h_{t}\{\srfc\}}\form,
\end{split}
\end{equation}
and we conclude that 
\begin{equation}
h_{t*}\curr_{\srfc}=\curr_{h_{t}\{\srfc\}},
\end{equation}
\ie, the image of the structure current induced by $\srfc$ is the
structure current induced by $h_{t}\{\srfc\}$. As expected, the defect
current satisfies
\begin{equation}
\bdry(h_{t*}\curr_{\srfc})=h_{t*}(\bdry\curr_{\srfc})=h_{t*}(\curr_{\bdry\srfc}).
\end{equation}

Next, we consider the rate of change of the pushforward of the structure
current and the associated defect current. One has, 
\begin{equation}
\begin{split}\resto{\parder[\tau]{}}{\tau=s}\left(\flow_{\tau,t*}\curr(\form)\right) & =\resto{\parder[\tau]{}}{\tau=s}\left(\curr(\flow_{\tau,t}^{*}\form)\right),\\
 & =\lim_{\Delta\tau\to0}\frac{\curr(\flow_{\tau+\Delta\tau,t}^{*}\form)-\curr(\flow_{\tau,t}^{*}\form)}{\Delta\tau},\\
 & =\lim_{\Delta\tau\to0}T\left(\frac{\flow_{\tau+\Delta\tau,t}^{*}\form-\flow_{\tau,t}^{*}\form}{\Delta\tau}\right).
\end{split}
\label{eq:RatesForCurrents-Computations}
\end{equation}
If 
\begin{equation}
\lim_{\Delta\tau\to0}\frac{\flow_{\tau+\Delta\tau,t}^{*}\form-\flow_{\tau,t}^{*}\form}{\Delta\tau},
\end{equation}
exists in the sense of test forms (not merely pointwise), then, one
may switch the order of the limit and the action of $\curr$ in the
last line of (\ref{eq:RatesForCurrents-Computations}) above. This
is indeed the case (see \cite[pp.~57--61]{derham}, and also \cite[pp.~132--135]{Giaquinta1998},
\cite[p.~363]{federer}). Hence, using (\ref{eq:flowAndLieDerivative}--\ref{eq:homotopyAndLieDerivative}),

\begin{equation}
\begin{split}\resto{\parder[\tau]{}}{\tau=s}\left(\flow_{\tau,t*}\curr(\form)\right) & =\curr\left(\resto{\parder[\tau]{}}{\tau=s}(\flow_{\tau,t}^{*}\form)\right),\\
 & =\curr(\flow_{s,t}^{*}(\lie_{\vf_{s}}\form)),\\
 & =(\flow_{s,t*}\curr)(\lie_{\vf_{s}}\form).
\end{split}
\end{equation}
It is noted that the Lie derivative operator on smooth forms with
compact supports is linear and continuous, so that one may define
its dual transformation $\lie_{\vf}^{*}$ on the space of currents
as
\begin{equation}
(\lie_{\vf}^{*}\curr)(\form)=\curr(\lie_{\vf}\form).
\end{equation}
It is also recalled that Cartan's magic formula for the Lie derivative
asserts that (\eg, \cite[p.~429]{AbeMarsdenManifolds})
\begin{equation}
\lie_{\vf}\form=\dee(\vf\contr\form)+\vf\contr\dee\form,
\end{equation}
so that
\begin{equation}
\begin{split}(\flow_{s,t*}\curr)(\lie_{\vf_{s}}\form) & =(\flow_{s,t*}\curr)(\dee(\vf_{s}\contr\form))+(\flow_{s,t*}\curr)(\vf_{s}\contr\dee\form),\\
 & =(\bdry\flow_{s,t*}\curr)(\vf_{s}\contr\form)+(\vf_{s}\wedge\flow_{s,t*}\curr)(\dee\form),\\
 & =\left(\vf_{s}\wedge(\bdry\flow_{s,t*}\curr)+\bdry(\vf_{s}\wedge\flow_{s,t*}\curr)\right)(\form).
\end{split}
\end{equation}
Thus, the rate of change of the structure current may be expressed
as
\begin{equation}
\resto{\parder[\tau]{}}{\tau=s}\flow_{\tau,t*}\curr=\lie_{\vf_{s}}^{*}\comp\flow_{s,t*}(T)=\vf_{s}\wedge(\bdry\flow_{s,t*}\curr)+\bdry(\vf_{s}\wedge\flow_{s,t*}\curr).
\end{equation}
In particular,
\begin{equation}
\resto{\parder[\tau]{}}{\tau=s}h_{\tau*}\curr=\lie_{\vf_{s}}^{*}\comp h_{s*}(T)=\vf_{s}\wedge(\bdry h_{s*}\curr)+\bdry(\vf_{s}\wedge h_{s*}\curr),
\end{equation}
and
\begin{equation}
\resto{\parder[\tau]{}}{\tau=t}\flow_{\tau,t*}\curr=\lie_{\vf_{t}}^{*}(T)=\vf_{t}\wedge\bdry\curr+\bdry(\vf_{s}\wedge\curr).\label{eq:RateOfCurr-Final}
\end{equation}

\subsection{Evolution of Smooth Distributions of Defects to Singular Ones}

The theory of currents provides a mathematical construction that may
be used to model the process at which a smooth distribution of defects
evolves and they coalesce into a ``macroscopic'' singular defect.

Similarly to Schwartz distributions, the action of general currents
can be approximated using currents induced by smooth forms through
the process of regularization or smoothing (see \cite[pp.~61--70]{derham},
\cite[pp.~346--348]{federer}, \cite[pp.~505--511]{Giaquinta1998}).
Specifically, given an $r$-current $\curr$, one can construct a
family of smooth $(n-r)$-forms $\gf_{\eps}$, $\eps\in(0,1]$ and
corresponding $\curr_{\eps}$ defined by 
\begin{equation}
\curr_{\eps}(\form)=\int_{\man}\gf_{\eps}\wedge\form,
\end{equation}
so that
\begin{equation}
\lim_{\eps\to0}\curr_{\eps}(\form)=\curr(\form).
\end{equation}
In other words, $\curr_{\eps}$ converge to $\curr$ weakly. Furthermore,
recalling that $\bdry T_{\eps}$ is the current induced by $\dee\gf_{\eps}$,
\ie, 
\begin{equation}
\bdry\curr_{\eps}(\psi)=(-1)^{n-r+1}\curr_{\dee\gf}(\psi)=(-1)^{n-r+1}\int_{\man}\dee\gf_{\eps}\wedge\psi,
\end{equation}
for each $(r-1)$-form $\psi$, the regularization process commutes
with the boundary operator so that 
\begin{equation}
\lim_{\eps\to0}\bdry\curr_{\eps}(\psi)=(-1)^{n-r+1}\lim_{\eps\to0}\int_{\man}\dee\gf_{\eps}\wedge\psi=\bdry\curr(\psi).
\end{equation}

Thus, setting $\curr_{0}=\curr$, and $t=1-\eps$, rather than a formal
mathematical approximation process, one could view the family $\curr_{t}$,
$t\in[0,1]$, as an evolution process of structure currents in the
time interval $[0,1]$ in which the smooth structure forms evolve
into a discrete structure current. Finally, the fact that smoothing
commutes with the boundary operator, implies that the smooth defect
forms evolve into the defect current.

\bigskip{}

\noindent \textbf{\textit{Acknowledgments.}} This work was partially
supported by Perlstone Center for Aeronautical Engineering Studies
at Ben-Gurion University.

\end{document}